\documentclass[english,12pt,aps,prd,a4paper,preprintnumbers,floatfix,nofootinbib,showpacs,superscriptaddress, notitlepage]{revtex4-1} 
\usepackage[mode=buildnew]{standalone}
\usepackage[usenames,dvipsnames]{color}  
\usepackage{graphicx}

\usepackage{setspace}
\usepackage{caption}
\usepackage{subcaption}
\usepackage{amsmath}
\usepackage{amssymb}
\usepackage[colorlinks=true,citecolor=darkred,urlcolor=darkred, pdfborder={0 0 0}]{hyperref}
\usepackage[normalem]{ulem}
\usepackage{xcolor}
\usepackage{placeins}
\usepackage{mathrsfs}
\usepackage{verbatim} 
\usepackage{cancel} 
\usepackage{float} 

\usepackage{scalerel}
\usepackage{tikz-feynman}
\tikzfeynmanset{compat=1.1.0}
\usepackage{feynmp}
\usepackage{tikzsymbols}
\usepackage{multirow}

\makeatletter
\def\p@subsection{}
\makeatother

\definecolor{darkred}{rgb}{0.6,0,0}

\definecolor{linkcolor}{rgb}{0,0,0.5}

\def\gsim{\raise0.3ex\hbox{$\;>$\kern-0.75em\raise-1.1ex\hbox{$\sim\;$}}}
\def\lsim{\raise0.3ex\hbox{$\;<$\kern-0.75em\raise-1.1ex\hbox{$\sim\;$}}}

\def\beqn#1{\begin{equation}\label{#1}}
\def\eeqn{\end{equation}}

\def\beqa#1{\begin{eqnarray}\label{#1}}
\def\eeqa{\end{eqnarray}}

\newcommand {\ignore}[1]{}

\def\UBL{$U(1)_{B+L}$ }
\def\Z4{$Z_4$}
\def\O5{$\mathcal{O}_5$ }

\def\321{$\mathrm{SU(3) \otimes SU(2) \otimes U(1)}$ }

\def\red{\color{red}{}}

\usepackage{orcidlink}
 
 \newcommand{\AddrIISERB}{Department of Physics,
 Indian Institute of Science Education and Research - Bhopal \\
 Bhopal Bypass Road, Bhauri, Bhopal, India}

\begin{document}

\title{\color{BrickRed} Dark Matter Induced Proton Decays}

\author{Ranjeet Kumar\orcidlink{0000-0002-7144-7606}}\email{ranjeet20@iiserb.ac.in}
\affiliation{\AddrIISERB}
\author{Rahul Srivastava\orcidlink{0000-0001-7023-5727}}\email{rahul@iiserb.ac.in}
\affiliation{\AddrIISERB}

\begin{abstract}
  \vspace{1cm} 

 We propose a novel theoretical framework in which proton decay is induced by the dark matter. While proton decay requires violation of the $B+L$ symmetry, dark matter stability often relies on the presence of an unbroken symmetry. These seemingly distinct phenomena are unified through the global \UBL symmetry inherent in the Standard Model. Its spontaneous breaking leads to a residual \Z4 symmetry, which ensures dark matter stability and forbids proton decay at tree level. Consequently, proton decay occurs at the one-loop level, mediated by dark sector particles. The proton lifetime is linked with the dark matter, the heavier dark matter mass enhancing proton stability, and vice versa.
The $\mathcal{O}$(TeV) masses of the mediators remain consistent with current proton lifetime limits, making them accessible to experimental searches.
In particular, the leptoquark mediating proton decay, carrying exotic $B+L$ charges, leads to a distinctive signature in collider searches. By intertwining proton decay, dark matter stability, and collider phenomenology,  this framework offers distinctive signatures that can be probed in current and future experiments.

\end{abstract}
\maketitle


\section{Introduction}
\label{sec:intro}

Within the Standard Model (SM), baryon number $(B)$ and lepton number $(L)$ are perturbatively conserved accidental global symmetries, which prevent processes like proton decay. However, this need not be true for higher dimensional operators. Indeed, the well-known dim-5 Weinberg operator~\cite{Weinberg:1979sa} violates lepton number by two units and generates the effective Majorana neutrino masses. At dim-6, considering only SM fields, there are five distinct flavor-independent operators that simultaneously break both $B$ and $L$ by one unit~\cite{Weinberg:1979sa,Wilczek:1979hc,Abbott:1980zj}.
These dim-6 operators directly enable two-body proton decays, providing potential decay channels that would be experimental signatures of baryon and lepton number violation. Crucially, such processes necessarily involve the violation of the combination $B+L$, highlighting that proton decay is not merely a consequence of breaking $B$ or $L$ individually, but a hallmark of $B+L$ violation. The significance of proton decay extends beyond particle physics, touching on fundamental cosmological questions. Therefore, while proton decay is yet to be observed, searching for it remains essential in probing high energy physics and unraveling the fundamental processes that allowed our matter-filled Universe to emerge from its early evolution.

Proton decay is generally discussed in the context of Grand Unified Theories (GUTs)~\cite{Pati:1973uk,Georgi:1974sy,Fritzsch:1974nn}. In GUTs, the proton is typically predicted to decay at the tree level. In many GUT models, the dominant decay modes proceed via two-body final states, with the associated new physics scale around $\Lambda \sim \mathcal{O}(10^{16})$ GeV. This scale is in agreement with the current limits set by the Super-Kamiokande (SK) experiment~\cite{Super-Kamiokande:2020wjk}, but it lies far beyond the reach of current and foreseeable collider experiments. As a result, the direct observation of proton decay remains elusive in collider experiments. 
 Despite extensive experimental searches, no direct evidence of proton decay has been observed, with only lower bounds established on the lifetime of various decay modes~\cite{Super-Kamiokande:2020wjk,Super-Kamiokande:2022egr}. Nevertheless, the next generation large volume detectors, namely DUNE~\cite{DUNE:2020ypp}, Hyper-Kamiokande~\cite{Hyper-Kamiokande:2018ofw}, and JUNO~\cite{JUNO:2021vlw} will offer a good chance to discover the baryon and lepton number violating processes.
However, this need not always be the case, as the masses of the mediators may lie within the reach of collider experiments, provided that the decay is induced at the loop level~\cite{Helo:2019yqp,Dorsner:2022twk,Nomura:2024zca,Kang:2024oyf}, or (and) through the higher dimensional operators~\cite{ODonnell:1993kdg,Babu:2012iv,Lehman:2014jma,Bhattacharya:2015vja,Liao:2016hru,Liao:2016qyd,Hambye:2017qix,Fonseca:2018ehk,Beneito:2023xbk,IBeneito:2025nby,Liao:2025vlj}.

Another key issue that remains unresolved in the SM is the absence of a viable dark matter (DM) candidate. Observational evidence, such as the results from the Planck satellite~\cite{Planck:2018vyg}, indicates that DM constitutes approximately 85\% of the matter content in the Universe, yet its fundamental nature remains elusive. The quest to identify DM particle(s) has led to numerous proposals, many of which also aim to resolve other unsolved problems within the SM. Among these, the scotogenic model~\cite{Ma:2006km} offers an innovative approach. In this framework, small neutrino masses are generated radiatively via loop processes involving dark sector particles, one of which serves as a stable DM candidate (see Refs.~\cite{Kang:2019sab,Leite:2020wjl,CentellesChulia:2022vpz,Batra:2022pej,Kumar:2023moh,Kumar:2024jot,Kumar:2024zfb,CentellesChulia:2024iom,Garnica:2024wur,Singh:2025jtn,Lozano:2025tst,Kumar:2025cte}). This dual purpose model addresses two of the most mysterious questions of the SM and provides a pathway for future experimental probes.
Building on this concept, we explore radiative proton decay, where the proton decays through one-loop diagrams mediated by dark sector particles. As we will argue in a later section, this process can effectively lower the new physics scale down to $\Lambda \sim \mathcal{O}(1)$ TeV, which can be probed in collider experiments contrary to the typical GUT models ($\Lambda \sim \mathcal{O}(10^{16})$ GeV). Similar to the scotogenic model~\cite{Ma:2006km}, the stability of the DM candidate is ensured by an unbroken discrete \Z4 symmetry.

We begin by revisiting the violation of the $B+L$ in the context of proton decay. In Ref.~\cite{Reig:2018yfd}, the authors proposed a framework for spontaneous proton decay via the breaking of axial \UBL symmetry to a discrete subgroup. Similarly, in our framework, we consider the breaking of \UBL symmetry to a discrete \Z4, which leads to proton decay. To accomplish this, we utilize the global \UBL symmetry, which is already present in the SM as an accidental symmetry. The spontaneous breaking of \UBL facilitates proton decay. This symmetry breaking gives rise to a residual \Z4 subgroup, which ensures DM stability and forbids proton decay at tree level. Proton decay is governed by a dim-6 operator, whose ultraviolet (UV)-completion involves dark sector particles at the one-loop level.
Within our framework, all dark sector particles carry odd charges under \Z4, while SM particles and the scalars $\chi$ and $\sigma$ remain even. Consequently, the lightest \Z4 odd particle serves as a stable and viable DM candidate.
The proton lifetime in this framework is controlled by the Yukawa couplings and the mass scale of the BSM particles running inside the loop. Interestingly, the same Yukawa interactions that govern the proton decay rate also affect the DM relic density and the DM-nucleon scattering cross section, due to direct interaction between dark sector particles and SM fermions. This interplay between proton decay and DM phenomenology imposes constraints on the model parameter space, allowing for a naturally suppressed proton decay rate while ensuring the stability of the DM candidate. In our model, BSM particles carrying exotic $B+L$ charges exhibit distinctive collider signatures. Altogether, this framework offers a compelling approach to addressing both proton decay and DM phenomenology within a parameter space that may be accessible to experiments.

The rest of the paper is organized as follows. In Sec.~\ref{sec:model}, we outline our model setup based on global \UBL symmetry and discuss its spontaneous breaking to the residual subgroup \Z4. In Sec.~\ref{sec:scalar}, we discuss the scalar sector and present the resulting mass spectrum. The fermion sector is discussed in Sec.~\ref{sec:ferm}. Sec.~\ref{sec:protondecay} provides a detailed discussion of the proton decay rate, along with numerical results demonstrating its dependence on the DM mass and the implications of breaking the residual \Z4 symmetry. The relic density and direct detection (DD) prospects of DM are presented in Sec.~\ref{sec:dm}. Collider searches and constraints on the leptoquark mass are discussed in Sec.~\ref{sec:coll}. Finally, we give concluding remarks in Sec.~\ref{sec:conc}.

\section{Model Framework} \label{sec:model}

We present a loop-induced proton decay framework in a scotogenic-like setup~\cite{Ma:2006km}, where the dark sector particles run in the loop and mediate proton decay. 
The stability of the DM candidate is ensured by the residual \Z4 arising from the global \UBL symmetry\footnote{Although we do not consider a gauged realization of the \UBL symmetry in the present work, for completeness, we outline the anomaly cancellation conditions in App.~\ref{sec:anomaly} that would need to be satisfied in a gauged \UBL scenario.}. We address proton decay in our model via the dim-6 effective operator $[d u][ue]$. To realize a UV-completion of this operator, we introduce new BSM particles: heavy charged leptons $(E_L,E_R)$, vector-like heavy EW singlet quarks $(U_L,U_R)$, and scalar leptoquark $\tilde{S}_1$. Additionally, a scalar $\zeta$ has been introduced, which is a singlet of the SM. Except for the Higgs field ($H$), all particles in the model carry non-trivial charges under the \UBL symmetry. This symmetry is spontaneously broken into a residual \Z4 symmetry through the vacuum expectation value (vev) of two additional scalar fields, $\chi$ and $\sigma$. We further introduce soft breaking terms to generate the mass for the would-be Goldstone boson that emerges from the spontaneous breaking of $U(1)_{B+L}$.
All SM particles are even under residual \Z4, while the dark sector BSM particles responsible for loop-induced proton decay are odd. The particle content and their transformations under `$\rm{SM} \otimes$\UBL' symmetry are summarized in Tab.~\ref{tab:particle}.
\begin{center}
 \begin{table}[!t]
\begin{tabular}{|c|| c || c | |c |}
  \hline
& Fields&  $SU(3)_C \otimes SU(2)_L \otimes U(1)_Y$  & $U(1)_{B+L}$   $\to$ {\red $Z_{4}$ }\ \\
\hline \hline
\ \multirow{6}{*}{\rotatebox{90}{SM }} \ & $Q_i$ \ \   & $(3, 2, \frac{1}{6})$ \ \   &  $\frac{1}{3}$ $\to$ {\red $\omega^{2}$} \\   
 &   $L_i$ \ \   & $(1, 2, -\frac{1}{2})$\ \  &  $1$  $\to$ {\red  $\omega^{2}$}  \\
& $u_i$ \ \   & $(3, 1, \frac{2}{3})$  \ \  &  $\frac{1}{3}$  $\to$ {\red   $\omega^{2}$}\\ 
& $e_i$  \ \  & $(1, 1, -1)$  \ \   &  $1$  $\to$ {\red  $\omega^{2}$}  \\
& $d_i$ \ \   & $(3, 1, -\frac{1}{3})$ \ \  &  $\frac{1}{3}$  $\to$ {\red  $\omega^{2}$} \\
&$H$\ \    & $(1, 2, \frac{1}{2})$  \ \   &  $0$    $\to$ {\red $1$} \\
\hline
\ \multirow{6}{*}{\rotatebox{90}{BSM }} \ &$(E_L,E_R)$  \ \   & $(1, 1, -1)$ \ \  &  $(\frac{5}{6},-\frac{1}{2})$  $\to$ {\red   $(\omega, \omega)$} \\
&$(U_L,U_R)$  \ \   & $(3, 1, \frac{2}{3})$ \ \  &  $(\frac{1}{2},-\frac{1}{6})$  $\to$ {\red    $(\omega^{3}, \omega^3)$ }\\
& $\tilde{S}_1$\ \   & $(\bar{3}, 1, \frac{4}{3})$ \ \  &  $\frac{1}{6}$  $\to$ {\red   $\omega$} \\
&$\sigma$ \ \   & $(1, 1, 0)$ \ \  &  $\frac{4}{3}$  $\to$ {\red   $1$} \\
&$\chi$ \ \   & $(1, 1, 0)$ \ \  &  $\frac{2}{3}$  $\to$ {\red   $1$ } \\
&$\boldsymbol{\zeta}$ \ \   & $\mathbf{(1, 1, 0)}$ \ \  &  $\mathbf{\frac{1}{6}}$  $\to$ { \red $\boldsymbol{ \omega}$ }\\    
\hline 
  \end{tabular}
\caption{The particle content and their transformations under various symmetries are presented for the case where the DM candidate is a scalar. An alternative UV-completion, featuring a fermionic DM candidate, is discussed in App.~\ref{sec:fer_dm}.}
  \label{tab:particle}
\end{table}
\end{center}
Notably, the scalar $\zeta$ is both color and electromagnetically neutral, making it a promising candidate for DM. While the UV-completion presented here yields a scalar DM candidate, our framework is flexible and allows for alternative realizations. For instance, by adjusting the particle charges and content, it is possible to construct a scenario where the DM candidate is a fermion, as we discuss in App.~\ref{sec:fer_dm}. 
Having discussed the particle content and their transformation properties, we next examine the scalar sector responsible for the symmetry breaking and mass generation of the particles.

\subsection{Scalar Sector} \label{sec:scalar}

\begin{figure}[h!]
     \centering
\includegraphics[width=0.45\linewidth]{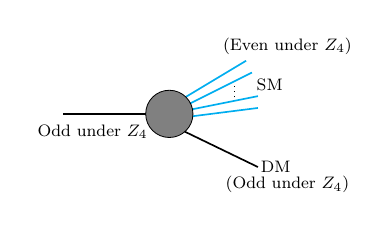}
\includegraphics[width=0.52\linewidth]{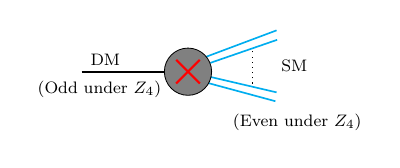}
     \caption{The diagrams depict the stability of the DM candidate under residual \Z4 symmetry. \textbf{Left:} the lightest odd sector particle is stable, \textbf{right:} the decay of DM candidate to SM particles is forbidden by the residual symmetry.}
     \label{fig:dm_stab}
 \end{figure}

We begin by discussing the scalar sector of our model, focusing on the mass spectrum of the scalar particles. This sector includes the familiar SM-like Higgs boson $H$ along with two additional SM singlet scalars, $\chi$ and $\sigma$, which drive the spontaneous breaking of the \UBL symmetry. In addition, there are two dark sector scalars, the leptoquark $\tilde{S}_1$ and the DM candidate $\zeta$. The complete scalar potential, invariant under the combined SM and \UBL symmetries, is given by\footnote{We note that in Eq.~\eqref{eq:pot}, soft-breaking cross terms between $\chi$ and $\sigma$ have been included within the last parentheses, ensuring that the would-be Goldstone boson acquires a mass. Importantly, these terms softly break the $U(1)_{B+L}$ symmetry while preserving the residual $Z_4$ symmetry.} 
\begin{small}
\begin{eqnarray} \label{eq:pot}
      V & = &  - \mu^2_{H} H^\dagger H -  \mu^2_{\chi} \chi^\dagger \chi- \mu^2_{\sigma}  \sigma^\dagger \sigma  + \mu^2_{\tilde{S}_1}  \tilde{S}_1^\dagger \tilde{S}_1 + \mu^2_{\zeta} \zeta^\dagger \zeta + \lambda_H \left( H^\dagger H \right)^2 +  \lambda_{\chi} \left( \chi^\dagger \chi \right)^2  \nonumber \\
   & + & \lambda_{\sigma} \left( \sigma^\dagger \sigma \right)^2+\lambda_{\tilde{S}_1} \left( \tilde{S}_1^\dagger \tilde{S}_1 \right)^2  + \lambda_{\zeta} \left( \zeta^\dagger \zeta \right)^2 + \lambda_{H \chi } \left( H^\dagger H \right) \left(\chi^\dagger \chi\right)  + \lambda_{H \sigma } \left( H^\dagger H \right) \left(\sigma^\dagger \sigma\right)  \nonumber \\
   &+ &  \lambda_{H \tilde{S}_1}  \left( H^\dagger H \right) \left(\tilde{S}_1^\dagger \tilde{S}_1\right) + \lambda_{H \zeta}  \left( H^\dagger H \right) \left(\zeta^\dagger \zeta\right)  + \lambda_{\chi \sigma } \left( \chi^\dagger \chi \right) \left( \sigma^\dagger \sigma \right) + \lambda_{\chi \tilde{S}_1 } \left( \chi^\dagger \chi \right) \left( \tilde{S}_1^\dagger \tilde{S}_1 \right)    \nonumber \\ 
  & + & \lambda_{\chi \zeta} \left( \chi^\dagger \chi \right) \left( \zeta^\dagger \zeta \right)+\lambda_{\sigma \tilde{S}_1} \left (\sigma^\dagger \sigma \right) \left( \tilde{S}_1^\dagger \tilde{S}_1 \right) + \lambda_{\sigma \zeta } \left (\sigma^\dagger \sigma \right) \left( \zeta^\dagger \zeta \right) + \lambda_{\tilde{S}_1 \zeta} \left (\tilde{S}_1^\dagger \tilde{S}_1 \right) \left( \zeta^\dagger \zeta \right)  \nonumber \\ 
  & + &    \left( \kappa \sigma^\dagger  \chi \chi + \rm{h.c.} \right) + \left(m^2_{\chi \sigma} \chi \sigma + m'^{2}_{\chi \sigma} \chi^\dagger  \sigma + \rm{h.c.}\right)  .
\end{eqnarray} 
\end{small}
Upon the symmetry breaking, the scalar fields $(H, \chi, \sigma)$ in the unitary gauge are given by
\begin{align} \label{eq:fieldexp}
    &H = \frac{1}{\sqrt{2}} \begin{pmatrix}                   
 0 \\
 v_H+\alpha_1
           \end{pmatrix}  ,\ \ \quad \chi=v_{\chi}+\alpha_2, \quad \sigma=v_{\sigma}+\alpha_3 \ .
\end{align}
The minimization conditions of the scalar potential can be derived as follows
\begin{align} \label{eq:tadpole}
    \mu^2_H&= \lambda_H v^2_H + \lambda_{ H \chi } v^2_{\chi}+ \lambda_{ H \sigma} v^2_{\sigma}, \nonumber \\
    \mu^2_{\chi} &= 2 \lambda_{\chi} v^2_{\chi} +\frac{1}{2} \lambda_{ H \chi}  v^2_H+\lambda_{\chi \sigma} v^2_{\sigma} + 2 \kappa v_{\sigma} + \left(m^{2}_{\chi \sigma} + m'^{2}_{\chi \sigma} \right) \frac{v_{\sigma}}{v_{\chi}},   \nonumber \\
   \mu^2_{\sigma} &= 2 \lambda_{\sigma} v^2_{\sigma} + \frac{1}{2}\lambda_{H\sigma}v^2_H+ \lambda_{\chi \sigma} v^2_{\chi} +  \frac{\kappa v^2_{\chi }}{v_{\sigma}} + \left(m^{2}_{\chi \sigma}  + m'^{2}_{\chi \sigma}  \right) \frac{v_{\chi}}{v_{\sigma}} \ . 
\end{align}
The fields $\alpha_1$, $\alpha_2$, $\alpha_3$ mix with each other, and their resulting mass squared matrix in the basis $(\alpha_1, \alpha_2, \alpha_3)^T$ is given by
\begin{align} \label{eq:higgsmass}
  \mathcal{M}^2_{\alpha}=  2 \begin{pmatrix}
         \lambda_H v^2_H &  \lambda_{H \chi} v_H v_{\chi} &  \lambda_{H \sigma} v_H v_{\sigma} \\
        (\ast) & 4 \lambda_{\chi} v^2_{\chi} - \left(m^{2}_{\chi \sigma} + m'^{2}_{\chi \sigma}  \right) \frac{v_{\sigma}}{v_{\chi}} & 2 \left( \lambda_{\chi \sigma}  v_{\sigma} + \kappa \right) v_{\chi} \\
        (\ast) & (\ast) & 4 \lambda_{\sigma} v^2_{\sigma} - \frac{ \kappa v^2_{\chi}}{v_{\sigma}} - \left(m^{2}_{\chi \sigma} + m'^{2}_{\chi \sigma}  \right) \frac{v_{\chi}}{v_{\sigma}} 
    \end{pmatrix} \ ,
\end{align}
where $(\ast)$ represents the symmetric part of the matrix.
The LHC data~\cite{CMS:2012qbp,ATLAS:2012yve,ParticleDataGroup:2024cfk} indicates that the mixing between the Higgs doublet $H$ and the singlet fields $(\chi, \sigma)$ is small. Thus, by taking the couplings $\lambda_{H \chi}$ and $\lambda_{H \sigma}$ to be small, we can simplify $\mathcal{M}_{\alpha}^2$, giving $m_h^2 \approx \lambda_H v^2_H = 125^2$ $\text{GeV}^2$.
The masses of DM candidate $\zeta$ and leptoquark $\tilde{S}_1$ are given by
\begin{align}
    m^2_{\zeta}&\equiv m^2_{\rm{DM}} = \mu^2_{\zeta} + \frac{1}{2}\lambda_{H \zeta} v^2_H +\lambda_{\chi \zeta} v^2_{\chi} +\lambda_{\sigma \zeta} v^2_{\sigma} \ . \label{eq:dmmass} \\
    M^2_{\tilde{S}_1}&= \mu^2_{\tilde{S}_1} + \frac{1}{2}\lambda_{H \tilde{S}_1} v^2_H +\lambda_{\chi \tilde{S}_1} v^2_{\chi} +\lambda_{\sigma \tilde{S}_1} v^2_{\sigma}\label{eq:lqmass} \ .
\end{align}
In our model, $\zeta$ being an SM singlet is identified as the DM candidate, protected by the residual $Z_4$ symmetry. This is illustrated by the diagrams presented in Fig.~\ref{fig:dm_stab}.  
 Under the residual \Z4 symmetry, all dark sector particles are odd, while SM particles are even. This imposes constraints on the possible decay modes of dark sector particles, ensuring that any such particle can only decay into SM particles, accompanied by another dark sector particle as depicted in the left panel of Fig.~\ref{fig:dm_stab}. Consequently, the lightest dark sector particle ($\zeta$) remains stable, making it a viable DM candidate as shown in the right panel of Fig.~\ref{fig:dm_stab}.
 The allowed mass spectrum of $\zeta$ and $\tilde{S}_1$ will have consequences on the proton lifetime and DM phenomenology and vice versa. The masses of DM candidate $\zeta$ and leptoquark $\tilde{S}_1$, as given in Eqs.~\eqref{eq:dmmass} and \eqref{eq:lqmass}, are determined not only by the corresponding mass parameters $\mu_{\zeta}$ and $\mu_{\tilde{S}_1}$, but also by the vev of the scalars, $H$, $\chi$, and $\sigma$. As a result, the same couplings that control DM annihilation processes also contribute to the mass of the DM candidate, distinguishing this scenario from the conventional singlet scalar DM framework~\cite{Cline:2013gha}. The vev of $\chi$ and $\sigma$ also influences the masses of the BSM fermions, as we discuss next.

\subsection{Fermion Sector} \label{sec:ferm}

In the fermion sector, we have two new BSM fermions in addition to the SM fermions, heavy charged lepton $(E_L, E_R)$ and vector-like quark $(U_L, U_R)$. Following the charge assignments given in Tab.~\ref{tab:particle}, the relevant BSM Yukawa Lagrangian for the model is given by
   \begin{align}
\label{eq:yuklag}
     -\mathcal{L}_{\rm{Yuk}} \supset   y_{\sigma} \overline{E_L} \sigma E_R  + y_{\chi} \overline{U_L} \chi U_R  + Y_1 \overline{U_L} \zeta u  + Y_2 U_R \tilde{S}_1^\dagger u + Y_3  E_R\tilde{S}_1 d+ Y_4  \overline{E_L} \zeta^{\dagger} e \ .
\end{align}
The first two terms of Eq.~\eqref{eq:yuklag} determine the mass generation of BSM fermions through the vev of $\sigma$ and $\chi$. The fermions $(E_L, E_R)$ and $(U_L, U_R)$ form Dirac pairs, and their corresponding mass eigenstates are denoted by $E$ and $U$, respectively. Their masses are given by the following expressions:
\begin{align}
   M_E=y_{\sigma} v_{\sigma} \ , \quad M_{U} = y_{\chi}v_{\chi} \ .
\end{align} 
Note that invariant mass terms for $E$ and $U$ are forbidden by the \UBL symmetry. The remaining Yukawa terms in Eq.~\eqref{eq:yuklag} influence proton decay, DM phenomenology, and collider searches. The Yukawa couplings $Y_i$ $(i=1,2,3,4)$ play a crucial role in shaping the proton decay rate, as they directly affect the one-loop processes within our model. In particular, for DM phenomenology, including relic density and WIMP-nucleon cross section, the Yukawa couplings $Y_1$ and $Y_4$ are especially important, as they govern the co-annihilation channels involving the scalar DM candidate $\zeta$. 
 
\section{Proton Decay} \label{sec:protondecay}

We now discuss the phenomenology of proton decay, which occurs at the one-loop level through the two-body process 
$p \rightarrow  e^{+} \pi^0$. The UV-complete Feynman diagram corresponding to this process is shown in Fig.~\ref{fig:pdfeyn}.
\begin{figure}[!h]
    \centering
    \begin{tikzpicture} 
\fill[black!100] (-3.2,0) ellipse (0.4 and 2.4);
\fill[black!100] (2.4,-1.8) ellipse (0.25 and 0.9);
\begin{feynman}
\vertex (cntf);
\vertex [above = 1cm of cntf] (cnt);
\vertex [above = 1.5cm of cnt] (sigv);
\vertex [below = 1.5cm of cnt] (chiv);
\vertex [above = 2.5cm of cnt] (sig);
\vertex [below = 2.5cm of cnt] (chi);
\vertex [above left = 1.5cm of cnt] (N1);
\vertex [below left = 1.5cm of cnt] (N31);
\vertex [above right = 1.5cm of cnt] (N2);
\vertex [below right = 1.5cm of cnt] (N32);
\vertex [below = 3cm of cnt] (fu);
\vertex [left=2cm of N1] (d);
\vertex [left=2cm of N31] (u2);
\vertex [left=3cm of fu] (u1);
\vertex [left=3.6cm of cntf] (p) {$p$};
\vertex [right=1.2cm of N2] (e) {$e^+$};
\vertex [right=1.2cm of N32] (u2rf);
\vertex [below = 1.2cm of u2rf] (u2r);
\vertex [below=1.0cm of u2r] (u1r);
\vertex [right = 2.6cm of chi] (pi) {$\pi^0$};

\diagram* {
(u1) -- [ fermion, edge label = $u$] (fu) -- [ fermion, edge label = $u$] (u1r),
(d) -- [ fermion, edge label = $d$] (N1),
(sigv) -- [anti fermion, edge label = $E_L$] (N2),
(sigv) -- [fermion, edge label' = $E_R$] (N1),
(e) -- [fermion] (N2),
(u2) -- [ fermion, edge label = $u$] (N31),
(N32) -- [anti fermion, edge label = $u$] (u2r),
(N31) -- [anti fermion, edge label' = $U_R$] (chiv) -- [anti fermion, edge label' = $U_L$] (N32),
(N1) -- [anti charged scalar, edge label' = $\tilde{S}_1$] (N31),
(N2) -- [ charged scalar, edge label = $\zeta$] (N32),
(sig) -- [ anti charged scalar, insertion = 0.01, edge label' = $\sigma$] (sigv),
(chi) -- [ anti charged scalar, insertion = 0.01, edge label' = $\chi$] (chiv),
};
\end{feynman}
\end{tikzpicture}
\caption{Proton decay at one-loop level mediated by new dark sector particles, where $\zeta$ is the DM candidate.}
\label{fig:pdfeyn}
\end{figure}
Due to the specific charge assignments of both SM and BSM particles under the residual \Z4 symmetry, proton decay is forbidden at tree level. The SM particles carry only even charges, while BSM particles have odd charges (except $\chi$ and $\sigma$) under \Z4. As a result, any tree level vertex facilitating proton decay through BSM particles would necessarily violate \Z4 symmetry, since the interaction would involve two even charges and one odd charge particle. Therefore, tree level proton decay is strictly forbidden by the residual \Z4 symmetry. Consequently, proton decay occurs at the loop level via a radiative process. The resulting decay rate can be significantly suppressed, allowing us to have mediator masses within the reach of current or near future colliders.

The width of proton decay induced at one-loop level can be expressed as
\begin{equation}
\Gamma\left(p \rightarrow  e^{+} \pi^0  \right)=\frac{m_p}{32 \pi}\left[1-\frac{m_{\pi^{0}}^2}{m_p^2}\right]^2\left|W_0 \mathcal{C}\right|^2,
\end{equation}
where $W_0$ is the hadronic matrix element, $W_0 \equiv -0.131$ $\rm{GeV}^2$~\cite{Aoki:2017puj} and $\mathcal{C}$ is the loop factor given by 
\begin{equation}
    \mathcal{C}=-|Y|^4 M_E M_U \mathcal{I}. 
\end{equation}
Here, $|Y|= \left(  Y_1 Y_2 Y_3 Y_4 \right)^{1/4}$, and $M_E$, $M_U$ are the masses of BSM fermions respectively, while $\mathcal{I}$ denotes the loop integral, the detailed derivation of which is provided in App.~\ref{app:proton}.

\begin{figure}[h!]
\centering
      \includegraphics[width=0.8\textwidth]{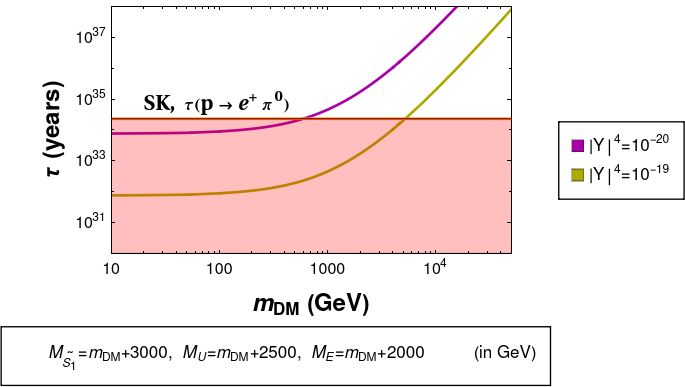}
    \caption{Correlation between the proton lifetime $\tau$ and the DM mass $m_{\rm{DM}}$. The horizontal black line indicates the SK limit on the proton lifetime~\cite{Super-Kamiokande:2020wjk}.}
    \label{fig:pd}
\end{figure}

We now present our model’s predictions for the proton decay rate and corresponding lifetime. In our framework, the most constraining proton decay process is the two-body mode $p \rightarrow  e^+ \pi^0$, which occurs at the one-loop level. The results for proton lifetime as a function of DM mass are shown in Fig.~\ref{fig:pd}.
The proton lifetime is primarily determined by the masses of the BSM particles running in the loop and Yukawa couplings. Notably, one of these particles is the DM candidate, establishing a direct correlation between the DM mass and proton lifetime. Heavy DM enhances proton stability, while a lighter DM mass reduces it. Conversely, the proton lifetime limit also constrains the viable parameter space for DM mass, as we discuss in Sec.~\ref{sec:dm}. The benchmark values are given in Fig.~\ref{fig:pd}, where we ensure that the DM candidate remains the lightest among all dark sector particles.
The proton lifetime predictions in our model are shown in magenta and yellow, corresponding to two different sets of Yukawa couplings. The black horizontal line shows the current lower limit on the proton lifetime provided from the SK experiment~\cite{Super-Kamiokande:2020wjk}, and the red shaded region remains ruled out from this limit. From Fig.~\ref{fig:pd}, it becomes evident that increasing the mediator particle masses results in a longer proton lifetime, bringing it into agreement with current experimental limits. This behavior is expected, as heavier mediators suppress loop-induced decay processes.

\subsection{Consequences of discrete \Z4 symmetry violation}

We now discuss the phenomenological implications of breaking the residual discrete \Z4 symmetry. In our model, \Z4 plays a crucial role: it ensures the stability of the DM candidate and forbids proton decay at tree level. However, if \Z4 is broken, both of these protections are lost, the DM candidate becomes unstable, and tree level proton decay becomes allowed within the framework. As a consequence, suppressing the proton decay rate to within experimental limits~\cite{Super-Kamiokande:2020wjk} requires the mediator particles to have masses near the GUT scale, placing them well beyond the reach of current collider experiments. 
\begin{figure}[!h]
\centering
\begin{tikzpicture}
\node (A) at (0,0) {\scalebox{0.8}{\begin{tikzpicture} 
\fill[black!100] (-3.2,0) ellipse (0.4 and 2.4);
\fill[black!100] (2.4,-1.8) ellipse (0.25 and 0.9);
\begin{feynman}
\vertex (cntf);
\vertex [above = 1cm of cntf] (cnt);
\vertex [above = 1.5cm of cnt] (sigv);
\vertex [below = 1.5cm of cnt] (chiv);
\vertex [above = 2.5cm of cnt] (sig);
\vertex [below = 2.5cm of cnt] (chi);
\vertex [above left = 1.5cm of cnt] (N1);
\vertex [below left = 1.5cm of cnt] (N31);
\vertex [above right = 1.5cm of cnt] (N2);
\vertex [below right = 1.5cm of cnt] (N32);
\vertex [below = 3cm of cnt] (fu);
\vertex [left=2cm of N1] (d);
\vertex [left=2cm of N31] (u2);
\vertex [left=3cm of fu] (u1);
\vertex [left=3.6cm of cntf] (p) {$p$};
\vertex [right=1.2cm of N2] (e) {$e^+$};
\vertex [right=1.2cm of N32] (u2rf);
\vertex [below = 1.2cm of u2rf] (u2r);
\vertex [below=1.0cm of u2r] (u1r);
\vertex [right = 2.6cm of chi] (pi) {$\pi^0$};

\diagram* {
(u1) -- [ fermion, edge label = $u$] (fu) -- [ fermion, edge label = $u$] (u1r),
(d) -- [ fermion, edge label = $d$] (N1),
(sigv) -- [anti fermion, edge label = $E_L$] (N2),
(sigv) -- [fermion, edge label' = $E_R$] (N1),
(e) -- [fermion] (N2),
(u2) -- [ fermion, edge label = $u$] (N31),
(N32) -- [anti fermion, edge label = $u$] (u2r),
(N31) -- [anti fermion, edge label' = $U_R$] (chiv) -- [anti fermion, edge label' = $U_L$] (N32),
(N1) -- [ anti charged scalar, edge label' = $\tilde{S}_1$] (N31),
(N2) -- [charged scalar, edge label = $\zeta$] (N32),
(sig) -- [anti charged scalar, insertion = 0.01, edge label' = $\sigma$] (sigv),
(chi) -- [anti charged scalar, insertion = 0.01, edge label' = $\chi$] (chiv),
};
\end{feynman}
\draw[->, thick, black] (3,0.5) -- (5.5,0.5); 
\node at (4.25,0.9){$\mathbf{Z_4}$\textbf{-breaking}};
\node at (0.0,-2.75){\textbf{Conserved} $\mathbf{Z_4}$};
\end{tikzpicture}}};
\node[below=0.3cm of A] {\small (a) Loop-induced proton decay conserving residual \Z4};

\node (B) at (8, 3) {\scalebox{0.8}{\begin{tikzpicture} 
\fill[black!100] (-3.2,0.4) ellipse (0.4 and 2.4);
\fill[black!100] (0.55,-0.750) ellipse (0.25 and 0.9);
\begin{feynman}
\vertex (cntf);
\vertex [above = 1cm of cntf] (cnt);
\vertex [right = 0.5cm of cnt] (ncnt);
\vertex [above = 1.5cm of ncnt] (sigv){$e^+$};
\vertex [below = 1.3cm of ncnt] (chiv);
\vertex [above = 2.5cm of cnt] (sig);
\vertex [below = 2.5cm of cnt] (chi);
\vertex [above left = 1.5cm of cnt] (N1);
\vertex [below left = 1.5cm of cnt] (N31);
\vertex [above right = 1.5cm of cnt] (N2);
\vertex [below right = 1.5cm of cnt] (N32);
\vertex [below = 2.2cm of ncnt] (fu);
\vertex [left=2cm of N1] (d);
\vertex [left=2cm of N31] (u2);
\vertex [left=3.5cm of fu] (u1);
\vertex [left=3.6cm of cntf] (p) {$p$};
\vertex [right=1.2cm of N2] (e);
\vertex [right=1.2cm of N32] (u2rf);
\vertex [below = 1.2cm of u2rf] (u2r);
\vertex [below=1.0cm of u2r] (u1r);

\diagram* {
(u1) -- [ fermion, edge label = $u$] (fu),
(d) -- [ fermion, edge label = $d$] (N1),
(sigv) -- [fermion] (N1),
(u2) -- [ fermion, edge label = $u$] (N31),
(N31) -- [anti fermion, edge label = $u$] (chiv), 
(N1) -- [ anti charged scalar, edge label' = $\tilde{S}_1$] (N31),
};
\end{feynman}
\node at (1.2,-0.8){$\pi^0$};
\node at (-1.0,-2.75){\textbf{Broken} $\mathbf{Z_4}$};
\end{tikzpicture}}};
\node[below=0.3cm of B] {\small (b) Tree level proton decay via $\tilde{S}_1$ violating \Z4};

\node (C) at (8, -4) {\scalebox{1.2}{

\tikzset{
    DM/.style={draw=black,line width=0.4pt, postaction={decorate}},
    E/.style={draw=black,line width=0.4pt, postaction={decorate}},
    scalar/.style={dotted,draw=black,line width=0.5pt, postaction={decorate},
        decoration={markings,mark=at position .55 with }},
    SMM/.style={draw=cyan,line width=0.6pt, postaction={decorate}},
    vertex/.style={draw=black, fill=black, circle, inner sep=0pt, minimum size=1mm}
}

%
%
%
%

\begin{tikzpicture}
\begin{feynman}
\vertex (a1) ;
\vertex[right=1.5cm of a1] (a2);
\vertex[right=1.0cm of a2] (a3) ;
\vertex[below=1.0cm of a3] (a4);
\vertex[above=1.0cm of a3] (a5);
\vertex[right=1.0cm of a5] (a6);
\vertex[below=1.0cm of a6] (a7);
\vertex[above=1.0cm of a6] (a8);

\diagram* {
(a1) -- [scalar,black] (a2),
(a4) -- [fermion, black] (a2),
(a2) -- [fermion, black] (a5),
};
\end{feynman}
\draw[fill=gray] (2.5, 0.8) ellipse (0.4 and 0.4);
\draw[scalar] (3.1, 1.35) -- (3.1, 0.96);
\draw[SMM] (2.7, 1.15) -- (3.3, 1.9);
\draw[SMM] (2.8, 1.05) -- (3.45, 1.68); 
\draw[SMM] (2.9, 0.9) -- (3.8, 0.75); 
\draw[SMM] (2.9, 0.75) -- (3.75, 0.52); 
\node at (0.65, 0.2) {\scriptsize DM};
 \node at (0.65, -0.2) {\scriptsize ( $Z_4$-odd )};
    \node at (3.5, 1.2) {\scriptsize SM};
    \node at (3.35, 2.10) {\scriptsize ( $Z_4$-even )};
    \node at (3.75, -1.25) {\scriptsize ( $Z_4$-even )};
    \node at (2.55, -1.25) {  $e^+/u$};
    \node at (2.2, 0.10) { \scriptsize $E/U$};
\end{tikzpicture}}};
\node[below=0.3cm of C] {\small (c) DM decay when \Z4 is broken};

\end{tikzpicture}

\caption{Representative Feynman diagrams illustrating loop-level (a, conserved \Z4) and tree level (b, broken \Z4) proton decay and DM decay (c, broken \Z4) mechanisms in our model.}
\label{fig:proton_decay_combined}
\end{figure}
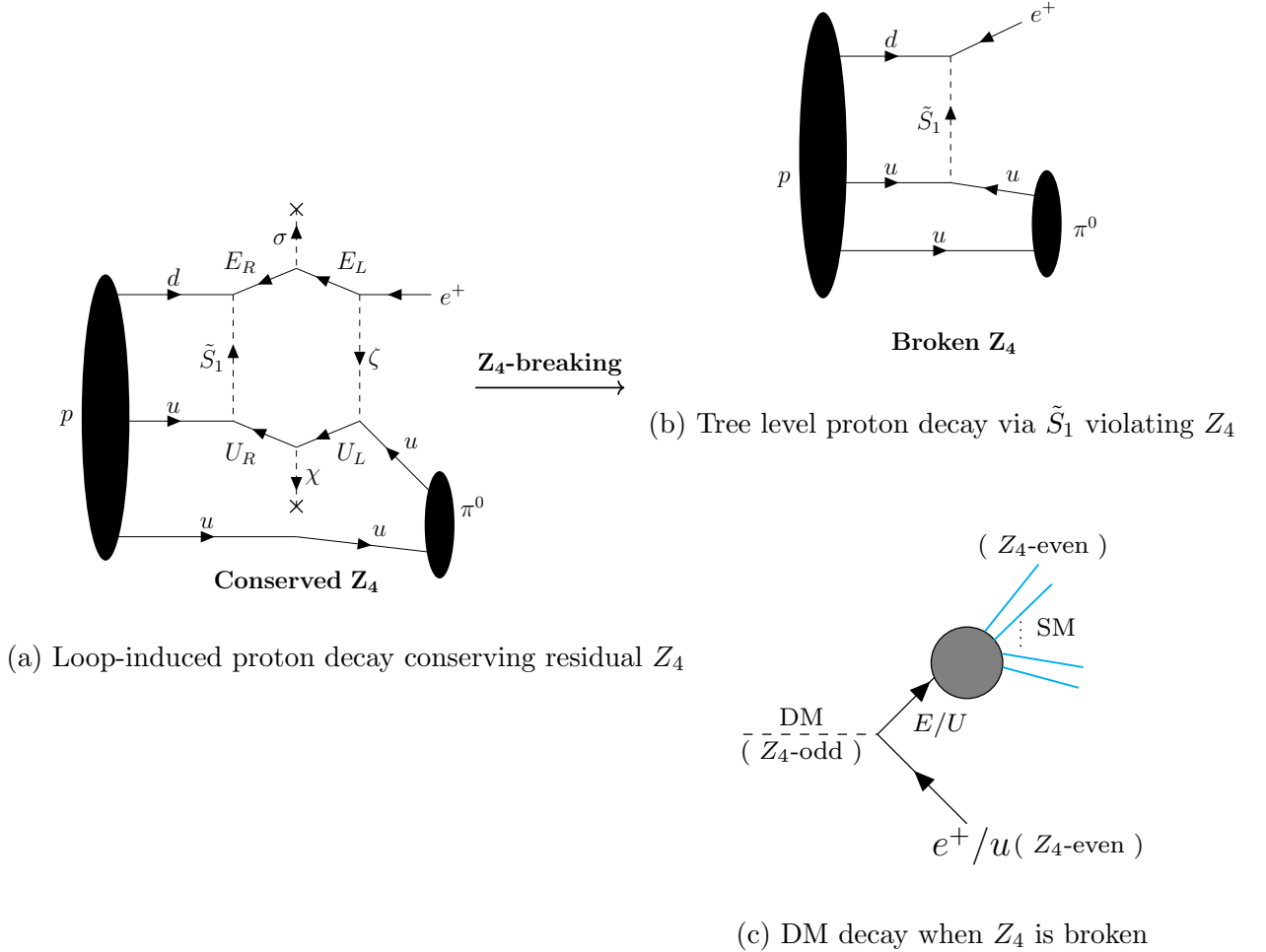
Specifically, once residual \Z4 is broken, the loop-induced process shown in Fig.~\ref{fig:proton_decay_combined}(a) becomes sub-leading, being superseded by the tree level diagram as depicted in Fig.~\ref{fig:proton_decay_combined}(b). Furthermore, the DM candidate is no longer stable with decay channels like Fig.~\ref{fig:proton_decay_combined}(c) opening up due to loss of the protective $Z_4$ symmetry.  

\section{Dark Matter} \label{sec:dm}
 
\begin{figure}[!h]
\centering
\includegraphics[width=0.78\textwidth]{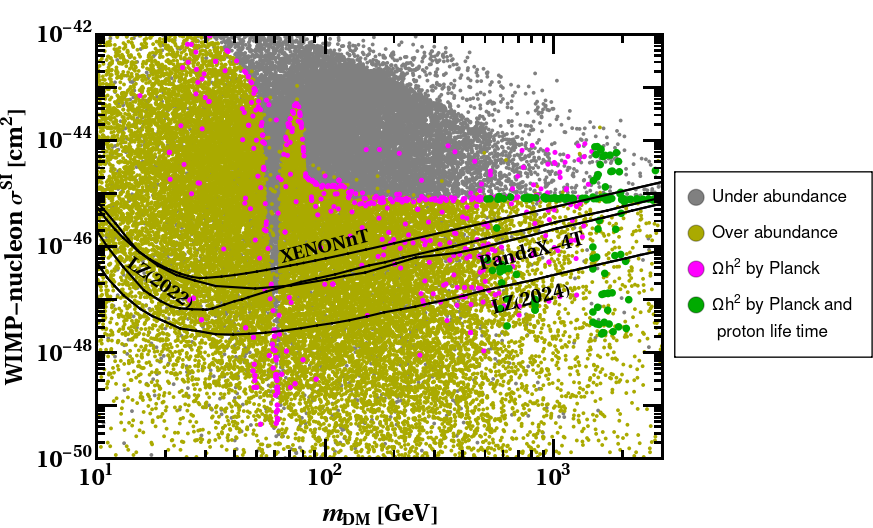}
    \caption{The variation of the spin-independent WIMP–nucleon cross section with respect to the DM mass $m_{\rm{DM}}$ is presented.}
    \label{fig:dm_dd}
\end{figure}

In the dark sector, we have two color neutral particles, scalar $\zeta$ and fermion $E$. In the present scenario, scalar $\zeta$ is also electromagnetically neutral, making it a viable DM candidate. However, it is possible to construct a scenario where the DM candidate is a fermion, as we discuss in App~\ref{sec:fer_dm}. 
\begin{figure}[!h]
\centering
      \includegraphics[width=0.60\textwidth]{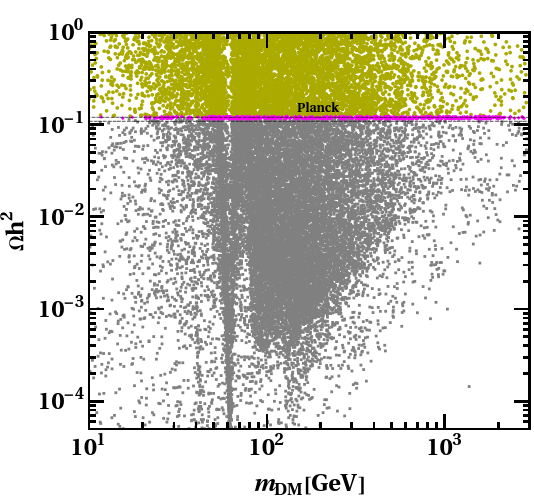}
    \caption{The relic density as a function of DM mass $m_{\rm{DM}}$ is shown. The gray, magenta, and yellow colors represent the under, correct, and over abundance of relic density, respectively.}
    \label{fig:dm_rel}
\end{figure}
We now present the numerical results for DM phenomenology. We start with the computation of the relic density for the DM candidate $\zeta$. Fig.~\ref{fig:dm_rel} depicts the plot of relic density as a function of the DM mass. The narrow band shown by black dotted lines corresponds to the $3 \sigma$ range for the DM relic density reported by the Planck satellite data~\cite{Planck:2018vyg},
\begin{align*}
0.1126 \leq \Omega_{\rm{DM}} h^{2} \leq 0.1246\;.
\end{align*}
The points shown in magenta correspond to the correct relic density points. The points shown in gray and yellow represent the under and over abundance of relic density, respectively. In the dark sector, in addition to the DM candidate, there are other BSM particles. While generating these points, we ensure that the DM candidate remains the lightest particle. The masses of the remaining dark sector particles are varied within the range $(1- 10^6)$ GeV. However, some of the resulting benchmark points are excluded by existing collider constraints. Only the green points in Fig.~\ref{fig:dm_dd} remain allowed, as we discuss below.

In our model, the DM candidate $\zeta$ is a singlet of $SU(2)_L$, therefore, only the Higgs portal serves as the primary annihilation channel for DM. A significant drop in the relic density is observed when the DM mass is approximately half the Higgs mass. At this point, the DM can annihilate efficiently via Higgs boson exchange. This results in an increased annihilation cross section, which consequently reduces the relic density of DM. Recently, in Ref.~\cite{Bharadwaj:2024crt}, it has been shown that in models where the Higgs portal dominates DM annihilation, loop corrections to the Higgs mass naturally impose an upper bound on the viable DM mass, typically around a few TeV. A similar constraint is expected in our scenario, however, a dedicated analysis is required due to additional complexities introduced by the mixing of the Higgs with the scalars $\chi$ and $\sigma$, as well as the presence of heavy fermions $E$ and $U$. Unlike the conventional singlet scalar DM model~\cite{Cline:2013gha}, our framework incorporates co-annihilation channels with the fermions $E$ and $U$. These additional channels allow the model to evade DD experimental constraints in the low mass range, as we discuss next.

 The DD prospects of DM candidate $\zeta$ are presented in Fig.~\ref{fig:dm_dd}, depicting the spin-independent DM-nucleon cross section as a function of the DM mass. We have also included co-annihilation channels in our analysis, leading to a wide range of viable relic density points, shown in magenta and green, that can evade current DD constraints~\cite{LZ:2022lsv,LZCollaboration:2024lux,XENON:2023cxc,PandaX:2024qfu}. Here, the spin-independent DM-nucleon cross section is mediated not only by the Higgs but also through direct couplings of the DM with quarks, as shown in Fig.~\ref{fig:pdfeyn}. Hence, the Yukawa couplings that govern the stability of the proton also play an important role in determining the relic density and cross section of the DM candidate. This interplay directly impacts the range of viable parameters, which are critical for both proton decay and DM phenomenology. The color code remains the same as in Fig.~\ref{fig:dm_rel}, except for the green color. These green points denote parameter regions consistent with the observed relic density~\cite{Planck:2018vyg}, current bounds on proton lifetime~\cite{Super-Kamiokande:2020wjk}, and existing collider constraints~\cite{ATLAS:2016wab,CMS:2017abv,ATLAS:2017mjy,CMS:2019zmd}. The current DD upper bounds from LZ (2022, 2024)~\cite{LZ:2022lsv,LZCollaboration:2024lux}, XENONnT~\cite{XENON:2023cxc}, and PandaX-4T~\cite{PandaX:2024qfu} have been shown as solid black lines. Among these, LZ (2024)~\cite{LZCollaboration:2024lux} provides the most stringent limit. Therefore, the green points lying below the LZ (2024) exclusion limit correspond to viable parameter space consistent with relic density, DD constraints, and the SK limits~\cite{Super-Kamiokande:2020wjk} on proton lifetime. Notably, these allowed points span a DM mass range from approximately 500 GeV to a few TeV. As a result, within our model framework, the requirement of proton stability can impose a  lower bound on the DM mass over the scanned parameter space. While conventional DM models often find the low mass region disfavored due to collider constraints (e.g., LEP and LHC) or DD limits, our framework offers a distinct perspective where such regions may evade these bounds. However, the proton lifetime constraint introduces an independent and complementary limitation that disfavors part of this otherwise viable low mass parameter space.

\section{Collider Signature} \label{sec:coll}
Apart from the DM candidate $\zeta$, our model features a rich BSM sector that includes the leptoquark $\tilde{S_1}$, a heavy charged lepton $E$, and a vector-like quark $U$. The searches for these particles are ongoing in the collider experiments. The conventional collider searches for these BSM particles do not directly apply to our model, as they carry exotic charges under $B+L$ symmetry and are also charged under the residual $Z_4$ symmetry. Consequently, their decay modes differ significantly from those considered in standard search strategies. 
To compute the cross sections and decay widths for the various processes, we first generated the corresponding UFO files of the model using the SARAH package~\cite{Staub:2015kfa}. These files were then imported into MadGraph5~\cite{Alwall:2014hca} for further numerical calculations.

\subsection{Collider Signature of Leptoquark $\tilde{S_1}$}

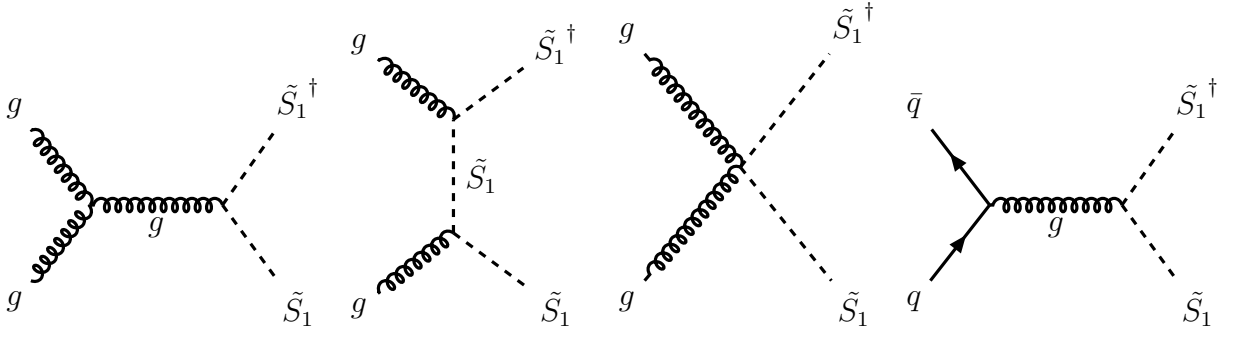
\begin{figure}[!h]
    \centering
    \begin{tikzpicture}
\begin{feynman}
\vertex (a1) ;
\vertex (a2) ;
\vertex (a3) ;
\vertex (i1) ;
\vertex[right=2.75cm of a2] (a4);
\vertex[right=1.0cm of a4] (i2);
\vertex[right=1.0cm of i1] (a2);
\vertex[below=1.0cm of i1] (a1) {\(g\)};
\vertex[above=1.0cm of i1] (a3) {\(g \)};
\vertex[below=1.0cm of i2] (a5){\( \tilde{S_1}\)};
\vertex[above=1.0cm of i2] (a6){\( \tilde{S_1}^{\dagger}\)};
\diagram* {
(a1) -- [gluon,black,very thick] (a2),
(a2) -- [gluon,black, very thick] (a3),
(a4) -- [gluon, black, very thick,edge label=\( g\)] (a2) ,
(a4) -- [scalar,black,very thick] (a6), 
(a4) -- [scalar,black,very thick] (a5), 
};
\end{feynman}
\end{tikzpicture}
\begin{tikzpicture}
\begin{feynman}
\vertex (a1) ;
\vertex (a2) ;
\vertex (a3) ;
\vertex (i) ;
\vertex[right=1.0cm of a1] (a2);
\vertex[left=1.0cm of i] (a1){\(g \)};
\vertex[right=1.0cm of a2] (a3);
\vertex[right=1.0cm of i] (a3){\(\tilde{S_1} \)};
\vertex[above=3.5cm of a1] (a5) {\(g \)};
\vertex[above=1cm of i] (a2) ;
\vertex[above=3.5cm of a3] (a6){\(\tilde{S_1}^{\dagger} \)};
\vertex[above=1.5cm of a2] (a4);
\diagram* {
(a2) -- [gluon,black,very thick] (a1),
(a2) -- [scalar,black, very thick] (a3),
(a2) -- [scalar, black, very thick, edge label'=\( \tilde{S_1} \)] (a4) ,
(a4) -- [gluon,black,very thick] (a5), 
(a4) -- [scalar,black, very thick ] (a6), 
};
\end{feynman}
\end{tikzpicture}
\vspace{0.10cm}
\vspace{0.10cm}
\begin{tikzpicture}
\begin{feynman}
\vertex (a1) ;
\vertex (a2) ;
\vertex (a3) ;
\vertex (i1) ;
\vertex[right=3.0cm of i1] (i2);
\vertex[right=1.5cm of i1] (a2);
\vertex[right=3.0cm of a1] (a5);
\vertex[right=3.0cm of a3] (a6);
\vertex[below=1.5cm of i1] (a1) {\(g \)};
\vertex[above=1.5cm of i1] (a3) {\(g \)};
\vertex[below=1.5cm of i2] (a5){\( \tilde{S_1}\)};
\vertex[above=1.5cm of i2] (a6){\( \tilde{S_1}^{\dagger}\)};
\diagram* {
(a2) -- [gluon,black,very thick] (a1),
(a2) -- [gluon,black, very thick] (a3),
(a2) -- [scalar,black,very thick] (a6), 
(a2) -- [scalar,black, very thick ] (a5), 
};
\end{feynman}
\end{tikzpicture}
\begin{tikzpicture}
\begin{feynman}
\vertex (a1) ;
\vertex (a2) ;
\vertex (a3) ;
\vertex (i1) ;
\vertex[right=2.75cm of a2] (a4);
\vertex[right=1.0cm of a4] (i2);
\vertex[right=1.0cm of i1] (a2);
\vertex[below=1.0cm of i1] (a1) {\(q\)};
\vertex[above=1.0cm of i1] (a3) {\(\bar{q} \)};
\vertex[below=1.0cm of i2] (a5){\( \tilde{S_1}\)};
\vertex[above=1.0cm of i2] (a6){\( \tilde{S_1}^{\dagger}\)};
\diagram* {
(a1) -- [fermion,black,very thick] (a2),
(a3) -- [anti fermion,black, very thick] (a2),
(a4) -- [gluon, black, very thick,edge label=\( g\)] (a2) ,
(a4) -- [scalar,black,very thick] (a6), 
(a4) -- [scalar,black,very thick] (a5), 
};
\end{feynman}
\end{tikzpicture}
\caption{Feynman diagrams depicting the pair production of leptoquark $\tilde{S}_1$ through gluon-gluon fusion and quark-antiquark annihilation.}
\label{fig:feyn_prodmodes1}
\end{figure}

\begin{figure}[h!]
     \centering
     \includegraphics[width=0.6
     \linewidth]{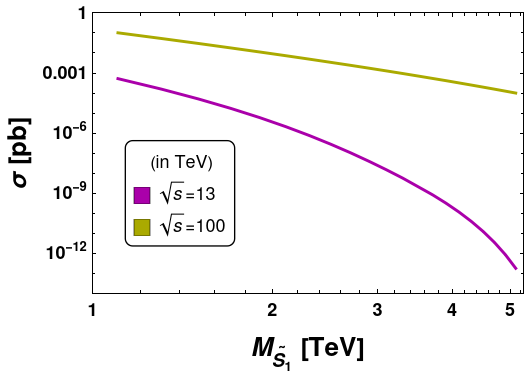}
     \caption{Production cross section of leptoquark $ \tilde{S}_1$ through pair production, $p p \to \tilde{S}_1 \tilde{S}_1^{\dagger}$, as a function of the mass of $\tilde{S}_1$, for the center-of-mass energies of $\sqrt{s}=13$ TeV and $\sqrt{s}=100$ TeV.}
     \label{fig:prod_s1t}
 \end{figure}

 We begin by first looking at signatures of the leptoquark $\tilde{S}_1$ in colliders. The leptoquarks have been extensively searched at LHC~\cite{Dorsner:2016wpm}. At the LHC, leptoquark pairs are primarily produced through gluon-gluon fusion and quark-antiquark annihilation, as shown in Fig.~\ref{fig:feyn_prodmodes1}. 
\begin{align}
    &g + g \to \tilde{S}_1 + \tilde{S}_1^{\dagger} \nonumber \\
     &q + \bar{q} \to \tilde{S}_1 + \tilde{S}_1^{\dagger}
\end{align}
These production channels remain the leading contributors to $\tilde{S}_1$ pair production in our scenario. In Fig.~\ref{fig:prod_s1t}, we show the corresponding production cross section of these processes at proton-proton (pp) colliders, for the LHC with $\sqrt{s}=13$ TeV, and for the Future Circular Collider-hadron (FCC-hh) with $\sqrt{s}=100$ TeV~\cite{FCC:2018vvp}.  
As expected, the cross section decreases rapidly with increasing mass due to phase space suppression at higher mass scales. At the LHC, the cross section drops below the femtobarn ($10^{-3}$ pb) level for $M_{\tilde{S}_1} \gtrsim 1$ TeV, making detection increasingly challenging. In contrast, the FCC-hh, with its significantly higher center-of-mass energy, can probe much heavier leptoquark masses with sizable cross sections. For instance, even at $M_{\tilde{S}_1} \sim 2$ TeV, the FCC-hh maintains cross sections above the femtobarn ($10^{-3}$ pb) level, offering promising discovery potential.
In addition, our model also allows for single leptoquark production. These production cross sections depend on the associated Yukawa couplings and are further suppressed by phase space. The sizes of these Yukawa couplings are constrained by the proton lifetime and the DM relic density. These couplings can be taken to be of $\mathcal{O}(1)$, while the remaining couplings and parameters are appropriately tuned to satisfy all relevant constraints. A detailed quantitative study of these effects would be interesting, but is beyond the scope of the present work.

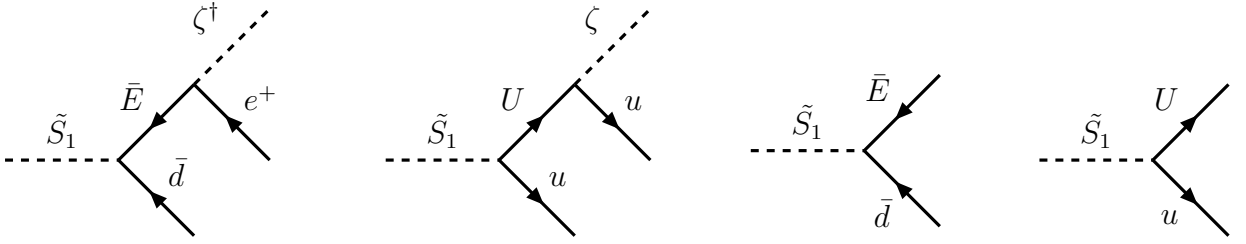
\begin{figure}[h!]
    \centering
    \begin{tikzpicture}
\begin{feynman}
\vertex (a1) ;
\vertex[right=1.5cm of a1] (a2);
\vertex[right=1.0cm of a2] (a3) ;
\vertex[below=1.0cm of a3] (a4);
\vertex[above=1.0cm of a3] (a5);
\vertex[right=1.0cm of a5] (a6);
\vertex[below=1.0cm of a6] (a7);
\vertex[above=1.0cm of a6] (a8);

\diagram* {
(a1) -- [scalar,black,very thick,edge label=\(\tilde{S_1}\)] (a2),
(a4) -- [fermion, black,very thick,edge label'=\(\bar{d}\)] (a2),
(a5) -- [fermion, black,very thick,edge label'=\(\bar{E}\)] (a2),
(a7) -- [fermion, black,very thick,edge label'=\(e^+\)] (a5),
(a5) -- [scalar, black,very thick,edge label=\(\zeta^{\dagger}\)] (a8),
};
\end{feynman}
\end{tikzpicture}
\hspace{1cm}
\begin{tikzpicture}
\begin{feynman}
\vertex (a1) ;
\vertex[right=1.5cm of a1] (a2);
\vertex[right=1.0cm of a2] (a3) ;
\vertex[below=1.0cm of a3] (a4);
\vertex[above=1.0cm of a3] (a5);
\vertex[right=1.0cm of a5] (a6);
\vertex[below=1.0cm of a6] (a7);
\vertex[above=1.0cm of a6] (a8);

\diagram* {
(a1) -- [scalar,black,very thick,edge label=\(\tilde{S_1}\)] (a2),
(a2) -- [fermion, black,very thick,edge label=\(u\)] (a4),
(a2) -- [fermion, black,very thick,edge label=\(U\)] (a5),
(a5) -- [fermion, black,very thick,edge label=\(u\)] (a7),
(a5) -- [scalar, black,very thick,edge label=\(\zeta\)] (a8),
};
\end{feynman}
\end{tikzpicture}
\hspace{1cm}
\begin{tikzpicture}
\begin{feynman}
\vertex (a1) ;
\vertex[right=1.5cm of a1] (a2);
\vertex[right=1.0cm of a2] (a3) ;
\vertex[below=1.0cm of a3] (a4);
\vertex[above=1.0cm of a3] (a5);
\vertex[right=1.0cm of a5] (a6);
\vertex[below=1.0cm of a6] (a7);
\vertex[above=1.0cm of a6] (a8);

\diagram* {
(a1) -- [scalar,black,very thick,edge label=\(\tilde{S_1}\)] (a2),
(a4) -- [fermion, black,very thick,edge label=\(\bar{d}\)] (a2),
(a5) -- [fermion, black,very thick,edge label'=\(\bar{E}\)] (a2),
};
\end{feynman}
\end{tikzpicture}
\hspace{1cm}
\begin{tikzpicture}
\begin{feynman}
\vertex (a1) ;
\vertex[right=1.5cm of a1] (a2);
\vertex[right=1.0cm of a2] (a3) ;
\vertex[below=1.0cm of a3] (a4);
\vertex[above=1.0cm of a3] (a5);
\vertex[right=1.0cm of a5] (a6);
\vertex[below=1.0cm of a6] (a7);
\vertex[above=1.0cm of a6] (a8);

\diagram* {
(a1) -- [scalar,black,very thick,edge label=\(\tilde{S_1}\)] (a2),
(a2) -- [fermion, black,very thick,edge label'=\(u\)] (a4),
(a2) -- [fermion, black,very thick,edge label=\(U\)] (a5),
};
\end{feynman}
\end{tikzpicture}
\caption{\centering Three-body and two-body decay modes of the leptoquark $\tilde{S_1}$.}
\label{fig:feyn_decaymodes1}
\end{figure}

Having analyzed the production of $ \tilde{S}_1$ at LHC and FCC-hh, we now turn our attention to its decay modes. The Feynman diagrams for different possible decay modes of  $\tilde{S}_1$ are shown in Fig.~\ref{fig:feyn_decaymodes1}. 
 \begin{figure}[h!]
     \centering
     \includegraphics[width=0.48\linewidth]{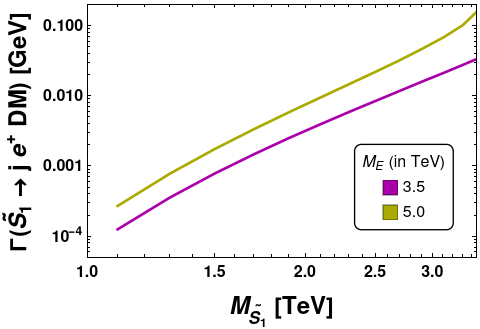}
\includegraphics[width=0.48\linewidth]{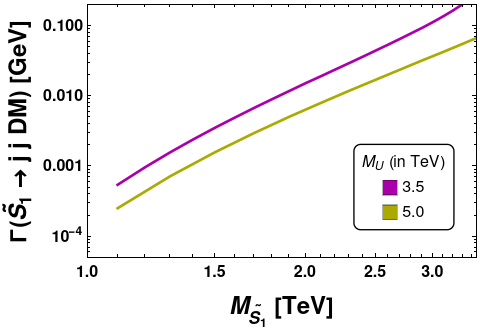}
    \caption{Three-body decay width of $\tilde{S}_1$ as a function of its mass, with the DM mass fixed at $m_{\rm{DM}}=1$ TeV.}
\label{fig:decaymodes_s1t1}
 \end{figure} 
The $ \tilde{S}_1$ decay channel depends critically on the mass hierarchy between $\tilde{S}_1$ and BSM fermions $E$ and $U$. If $ \tilde{S}_1$ is lighter than either $E$ or $U$, i.e., $M_{\tilde{S_1}}< M_E$ or $M_{\tilde{S_1}}<M_{U}$, its decay proceeds via three-body channels. These lead to final states such as $j e^+ \cancel{E}$ or $j j \cancel{E}$, where $j$ denotes jets and $\cancel{E}$ represents missing energy.
The corresponding decay widths for these three-body decay modes are shown in Fig.~\ref{fig:decaymodes_s1t1}. For Fig.~\ref{fig:decaymodes_s1t1}, we choose two benchmark values for the masses of BSM fermions: $M_E=(3.5,5.0)$ TeV and $M_U=(3.5,5.0)$ TeV, while fixing the DM mass at $m_{\rm{DM}}=1$ TeV. This choice is consistent with both the proton lifetime limit and DM constraints discussed in previous sections. The relevant Yukawa couplings are taken to be of order $\mathcal{O}(1)$, again respecting all the experimental constraints.
\begin{figure}[!h]
     \centering
\includegraphics[width=0.47\linewidth]{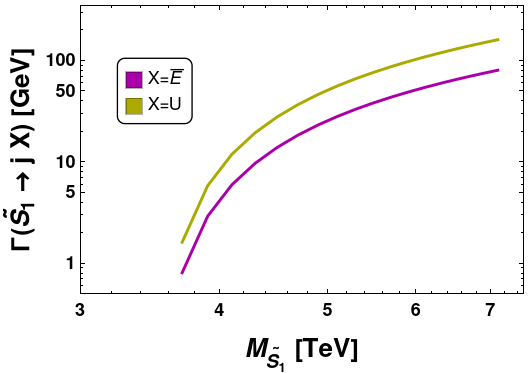}
\includegraphics[width=0.49\linewidth]{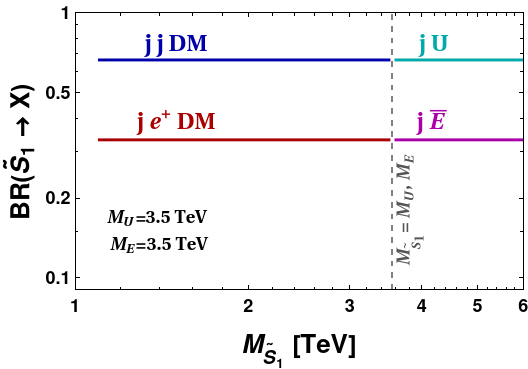}
     \caption{\textbf{Left:} two-body decay width of $\tilde{S}_1$ as a function of its mass. \textbf{Right:} branching ratios of $\tilde{S_1}$ for three-body and two-body decays as a function of its mass, with the DM mass fixed at $m_{\rm{DM}}=1$ TeV.}
     \label{fig:decaymodes_s1t2}
 \end{figure} 

On the other hand, if the leptoquark is heavier than BSM fermions, i.e., $M_{\tilde{S_1}}> M_E$ (and/or $M_{\tilde{S_1}}> M_{U}$), it decays predominantly through two-body modes: $\tilde{S_1} \to j X$ ($X=\bar{E}$ and/or $X= U$) as shown in the left panel of Fig.~\ref{fig:decaymodes_s1t2}. The masses of $E$ and $U$ are fixed at $3.5$ TeV. These decays are typically prompt for $\mathcal{O}(1)$ Yukawa couplings. From Fig.~\ref{fig:decaymodes_s1t1} and the left panel of Fig.~\ref{fig:decaymodes_s1t2}, it is evident that the two-body decay width of ${\tilde{S_1}}$ is considerably larger than the three-body decay width.

Since the decay modes of $\tilde{S_1}$, whether through three-body or two-body processes, depend on the masses of the BSM fermions ($E$ and $U$), we choose a benchmark value to illustrate this behavior. 
In the right panel of Fig.~\ref{fig:decaymodes_s1t2}, we present the branching ratios (BR) of these decay channels for $(M_E, M_U)=(3.5, 3.5)$ TeV. For  $M_{\tilde{S_1}} \lesssim 3.5$ TeV, the three-body decay mode dominates, whereas for $M_{\tilde{S_1}} > 3.5$ TeV, the two-body decay becomes the dominant channel.
 Although $\tilde{S_1}$ resembles a scalar leptoquark based on its charge assignments, standard leptoquark searches do not directly apply. Its decay modes differ from conventional leptoquark decays due to its non-trivial charge under the $B+L$ and the residual $Z_4$ symmetry. Nevertheless, its mass can be constrained using combined ATLAS leptoquark searches~\cite{ATLAS:2016wab,ATLAS:2017mjy} and SUSY searches~\cite{CMS:2017abv,CMS:2019zmd}, yielding a lower bound of $M_{\tilde{S_1}} \gtrsim \mathcal{O}(1)$ TeV. 

 \subsection{Collider Signature of heavy charged lepton $E$ and vector-like quark $U$} \label{sec:coll_UE}
\begin{figure}[!t]
    \centering
    \begin{tikzpicture}
\begin{feynman}
\vertex (a1) ;
\vertex (a2) ;
\vertex (a3) ;
\vertex (i) ;
\vertex[right=1.0cm of a1] (a2);
\vertex[left=1.0cm of i] (a1){\(q \)};
\vertex[right=1.0cm of a2] (a3);
\vertex[right=1.0cm of i] (a3){\(E \)};
\vertex[above=3.5cm of a1] (a5) {\(\bar{q} \)};
\vertex[above=1cm of i] (a2) ;
\vertex[above=3.5cm of a3] (a6){\(\bar{E} \)};
\vertex[above=1.5cm of a2] (a4);
\diagram* {
(a2) -- [anti fermion,black,very thick] (a1),
(a2) -- [scalar,black, very thick] (a3),
(a2) -- [scalar, black, very thick, edge label'=\( \tilde{S_1} \)] (a4) ,
(a4) -- [fermion,black,very thick] (a5), 
(a4) -- [scalar,black, very thick ] (a6), 
};
\end{feynman}
\end{tikzpicture}
\vspace{0.10cm}
\begin{tikzpicture}
\begin{feynman}
\vertex (a1) ;
\vertex (a2) ;
\vertex (a3) ;
\vertex (i1) ;
\vertex[right=2.75cm of a2] (a4);
\vertex[right=1.0cm of a4] (i2);
\vertex[right=1.0cm of i1] (a2);
\vertex[below=1.0cm of i1] (a1) {\(g\)};
\vertex[above=1.0cm of i1] (a3) {\(g \)};
\vertex[below=1.0cm of i2] (a5){\( U\)};
\vertex[above=1.0cm of i2] (a6){\( \bar{U}\)};
\diagram* {
(a1) -- [gluon,black,very thick] (a2),
(a2) -- [gluon,black, very thick] (a3),
(a4) -- [gluon, black, very thick,edge label=\( g\)] (a2) ,
(a4) -- [scalar,black,very thick] (a6), 
(a4) -- [scalar,black,very thick] (a5), 
};
\end{feynman}
\end{tikzpicture}
\begin{tikzpicture}
\begin{feynman}
\vertex (a1) ;
\vertex (a2) ;
\vertex (a3) ;
\vertex (i) ;
\vertex[right=1.0cm of a1] (a2);
\vertex[left=1.0cm of i] (a1){\(g \)};
\vertex[right=1.0cm of a2] (a3);
\vertex[right=1.0cm of i] (a3){\(U \)};
\vertex[above=3.5cm of a1] (a5) {\(g \)};
\vertex[above=1cm of i] (a2) ;
\vertex[above=3.5cm of a3] (a6){\(\bar{U} \)};
\vertex[above=1.5cm of a2] (a4);
\diagram* {
(a2) -- [gluon,black,very thick] (a1),
(a2) -- [scalar,black, very thick] (a3),
(a2) -- [scalar, black, very thick, edge label'=\( U \)] (a4) ,
(a4) -- [gluon,black,very thick] (a5), 
(a4) -- [scalar,black, very thick ] (a6), 
};
\end{feynman}
\end{tikzpicture}
\vspace{0.10cm}
\begin{tikzpicture}
\begin{feynman}
\vertex (a1) ;
\vertex (a2) ;
\vertex (a3) ;
\vertex (i1) ;
\vertex[right=2.75cm of a2] (a4);
\vertex[right=1.0cm of a4] (i2);
\vertex[right=1.0cm of i1] (a2);
\vertex[below=1.0cm of i1] (a1) {\(q\)};
\vertex[above=1.0cm of i1] (a3) {\(\bar{q} \)};
\vertex[below=1.0cm of i2] (a5){\( U\)};
\vertex[above=1.0cm of i2] (a6){\( \bar{U}\)};
\diagram* {
(a1) -- [fermion,black,very thick] (a2),
(a3) -- [anti fermion,black, very thick] (a2),
(a4) -- [gluon, black, very thick,edge label=\( g\)] (a2) ,
(a4) -- [scalar,black,very thick] (a6), 
(a4) -- [scalar,black,very thick] (a5), 
};
\end{feynman}
\end{tikzpicture}
\caption{Leading-order Feynman diagrams for the pair production of the heavy charged lepton $E$ and the vector-like quark $U$.}
\label{fig:feyn_prodmodes2}
\end{figure}
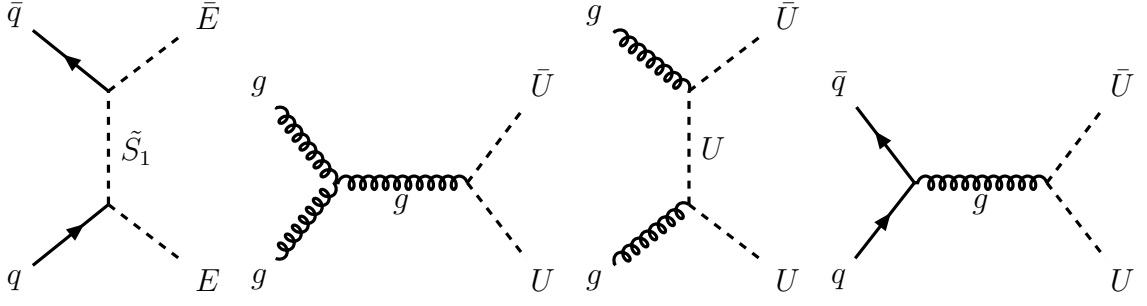

\begin{figure}[h!]
     \centering
      \includegraphics[width=0.48\linewidth]{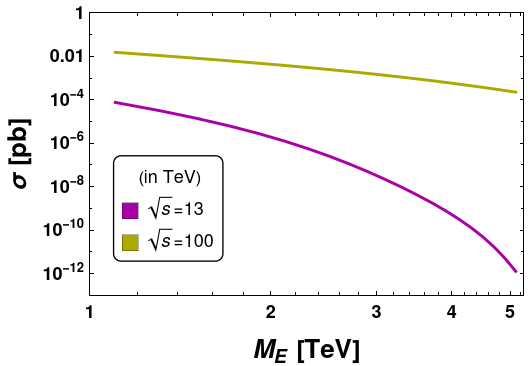}
    \includegraphics[width=0.48\linewidth]{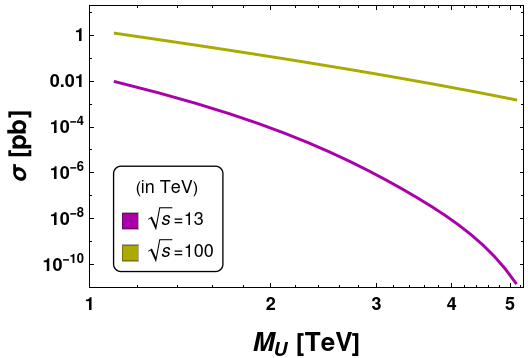}   
     \caption{Production cross section of the heavy charged lepton $E$ (left panels) and the vector-like quark $U$ (right panels) as functions of their masses, for the center-of-mass energies of $\sqrt{s}=13$ TeV and $\sqrt{s}=100$ TeV.}
     \label{fig:prod2}
 \end{figure}

 The heavy charged lepton $E$ and vector-like quark $U$ can be produced at hadron colliders through quark-antiquark annihilation. Additionally, since $U$ is a color-triplet, it can also be produced via gluon-gluon fusion, significantly enhancing its production rate due to strong interactions. The relevant Feynman diagrams illustrating these production mechanisms are shown in Fig.~\ref{fig:feyn_prodmodes2}. 
 The production cross section of these fermions at the LHC ($\sqrt{s}=13$ TeV) and FCC-hh ($\sqrt{s}=100$ TeV) has been shown in Fig.~\ref{fig:prod2}.
The production cross section for the fermion $E$ is significantly smaller than that for the fermion $U$. This difference arises because, in the case of $E$ pair production, there is no contribution from gluon-gluon fusion, which is a dominant production channel at hadron colliders. In contrast, the production of the $U$ benefits from strong interactions, leading to a much larger cross section. This effect is evident in Fig.~\ref{fig:prod2}, which shows the production rates of $E$ and $U$ at $\sqrt{s}=13$ TeV and $\sqrt{s}=100$ TeV.
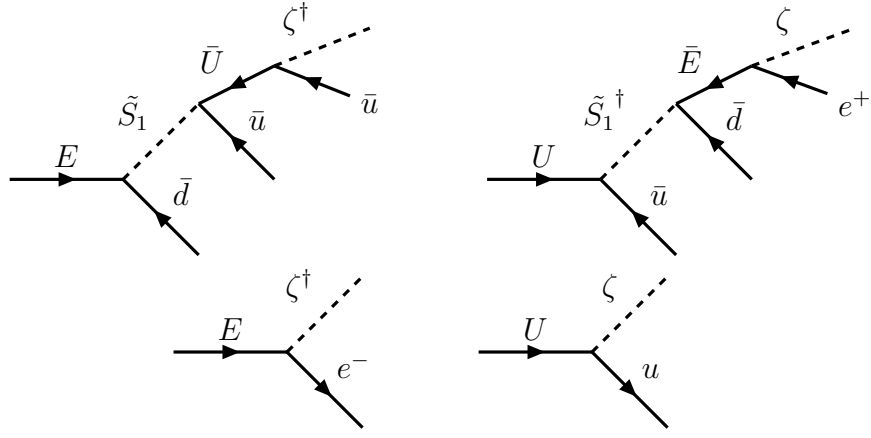
\begin{figure}[h!]
    \centering
    \hspace{0.5cm}
\begin{tikzpicture}
\begin{feynman}
\vertex (a1) ;
\vertex[right=1.5cm of a1] (a2);
\vertex[right=1.0cm of a2] (a3) ;
\vertex[below=1.0cm of a3] (a4);
\vertex[above=1.0cm of a3] (a5);
\vertex[right=1.0cm of a5] (a6);
\vertex[below=1.0cm of a6] (a7);
\vertex[above=0.5cm of a6] (a8);
\vertex[right=1.0cm of a6] (a9) {$\bar{u}$};
\vertex[above=1.0cm of a9] (a10) ;

\diagram* {
(a1) -- [fermion,black,very thick,edge label=\({E}\)] (a2),
(a4) -- [fermion, black,very thick,edge label'=\(\bar{d}\)] (a2),
(a2) -- [scalar, black,very thick,edge label=\(\tilde{S_1}\)] (a5),
(a7) -- [fermion, black,very thick,edge label'=\(\bar{u}\)] (a5),
(a8) -- [fermion, black,very thick,edge label'=\(\bar{U}\)] (a5),
(a9) -- [fermion, black,very thick] (a8),
(a8) -- [scalar, black,very thick,edge label=\(\zeta^{\dagger}\)] (a10),
};
\end{feynman}
\end{tikzpicture}
\hspace{1cm}
\begin{tikzpicture}
\begin{feynman}
\vertex (a1) ;
\vertex[right=1.5cm of a1] (a2);
\vertex[right=1.0cm of a2] (a3) ;
\vertex[below=1.0cm of a3] (a4);
\vertex[above=1.0cm of a3] (a5);
\vertex[right=1.0cm of a5] (a6);
\vertex[below=1.0cm of a6] (a7);
\vertex[above=0.5cm of a6] (a8);
\vertex[right=1.0cm of a6] (a9) {$e^+$};
\vertex[above=1.0cm of a9] (a10) ;

\diagram* {
(a1) -- [fermion,black,very thick,edge label=\({U}\)] (a2),
(a4) -- [fermion, black,very thick,edge label'=\(\bar{u}\)] (a2),
(a2) -- [scalar, black,very thick,edge label=\(\tilde{S_1}^\dagger\)] (a5),
(a7) -- [fermion, black,very thick,edge label'=\(\bar{d}\)] (a5),
(a8) -- [fermion, black,very thick,edge label'=\(\bar{E}\)] (a5),
(a9) -- [fermion, black,very thick] (a8),
(a8) -- [scalar, black,very thick,edge label=\(\zeta\)] (a10),
};
\end{feynman}
\end{tikzpicture}
\hspace{5cm}
\begin{tikzpicture}
\begin{feynman}
\vertex (a1) ;
\vertex[right=1.5cm of a1] (a2);
\vertex[right=1.0cm of a2] (a3) ;
\vertex[below=1.0cm of a3] (a4);
\vertex[above=1.0cm of a3] (a5);
\vertex[right=1.0cm of a5] (a6);
\vertex[below=1.0cm of a6] (a7);
\vertex[above=1.0cm of a6] (a8);

\diagram* {
(a1) -- [fermion,black,very thick,edge label=\(E\)] (a2),
(a2) -- [fermion, black,very thick,edge label=\(e^-\)] (a4),
(a2) -- [scalar, black,very thick,edge label=\(\zeta^\dagger\)] (a5),
};
\end{feynman}
\end{tikzpicture}
\hspace{1cm}
\begin{tikzpicture}
\begin{feynman}
\vertex (a1) ;
\vertex[right=1.5cm of a1] (a2);
\vertex[right=1.0cm of a2] (a3) ;
\vertex[below=1.0cm of a3] (a4);
\vertex[above=1.0cm of a3] (a5);
\vertex[right=1.0cm of a5] (a6);
\vertex[below=1.0cm of a6] (a7);
\vertex[above=1.0cm of a6] (a8);

\diagram* {
(a1) -- [fermion,black,very thick,edge label=\(U\)] (a2),
(a2) -- [fermion, black,very thick,edge label=\(u\)] (a4),
(a2) -- [scalar, black,very thick,edge label=\(\zeta\)] (a5),
};
\end{feynman}
\end{tikzpicture}
\caption{Four-body and two-body decay modes of the heavy charged lepton $E$ and the vector-like quark $U$.}
\label{fig:feyn_decaymodes2}
\end{figure}

We now discuss the decay modes of $E$ and $U$. These particles can undergo both four-body and two-body decays, as shown in Fig.~\ref{fig:feyn_decaymodes2}.
\begin{figure}[h!]
     \centering
     \includegraphics[width=0.48\linewidth]{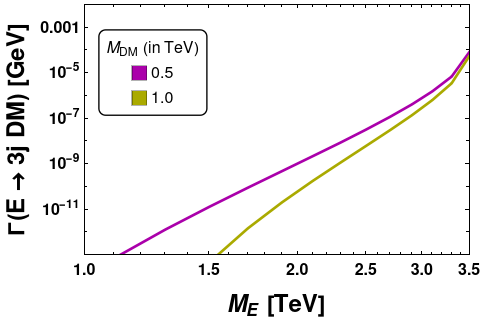}
\includegraphics[width=0.48\linewidth]{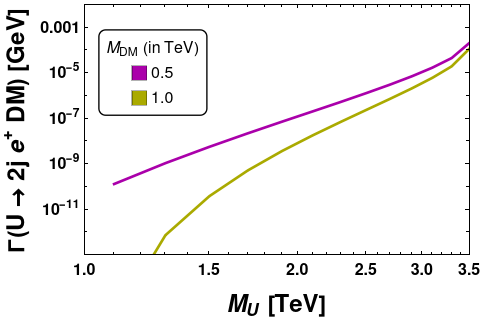}
\includegraphics[width=0.48\linewidth]{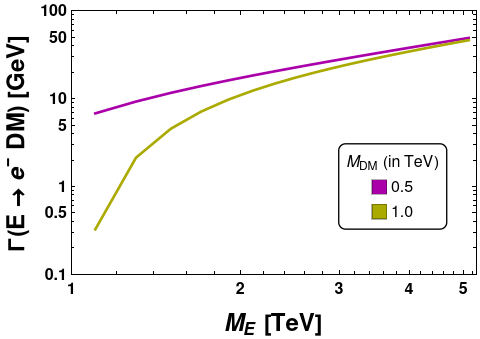}
\includegraphics[width=0.48\linewidth]{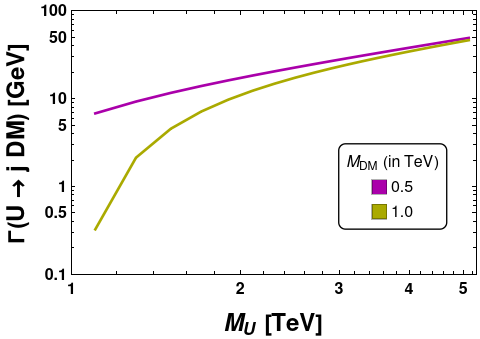}
     \caption{Decay width of $E$ (left panels) and $U$ (right panels) as a function of their masses for the four-body decay modes (top) and two-body decay modes (bottom).}
     \label{fig:decaymodes_ue}
 \end{figure}
We present the decay widths for both four-body and two-body decay modes in Fig.~\ref{fig:decaymodes_ue}, considering $\mathcal{O}(1)$ Yukawa couplings. The four-body decay of $E$ ($U$) is mediated by the heavy leptoquark state $\tilde{S_1}$ and the other BSM fermion $U$ ($E$), leading to a suppressed decay width due to the presence of multiple heavy intermediate states. In contrast, the two-body decay proceeds promptly and dominates, as it does not rely on any intermediate particles. This difference in decay rates can play a key role in determining the experimental signatures of these particles at colliders.

\section{Conclusions} \label{sec:conc}

In this work, we presented a theoretical framework for DM induced proton decay. This scenario arose naturally from the $U(1)_{B+L}$ symmetry, which is already present in the SM. Its spontaneous breaking through the vev of the scalar fields $\chi$ and $\sigma$ results in a residual \Z4 symmetry. This residual symmetry serves a dual purpose: it ensures the stability of the DM candidate and simultaneously forbids proton decay at tree level. As a result, proton decay occurs at the one-loop level via a dim-6 operator, which is UV-completed by the dark sector particles. Within this framework, all BSM particles (except $\chi$ and $\sigma$) are odd under the $Z_4$ symmetry, while SM particles remain even.

Notably, the dark sector particles exhibit direct coupling to SM fermions as shown in Fig.~\ref{fig:pdfeyn}, thereby influencing both the proton decay rate and the DM–nucleon scattering cross section. This interplay establishes a direct connection between the proton lifetime and the DM mass, wherein a heavier DM mass tends to enhance proton stability, and vice versa. Consequently, the combined constraints from proton lifetime, relic density, and DD experiments impose bounds on the DM mass as shown in Fig.~\ref{fig:dm_dd}. 
Furthermore, the mediators inducing proton decay have masses at the scale $\Lambda \sim \mathcal{O}(1)$ TeV, potentially accessible at current or near future collider experiments. Their exotic $B+L$ charges give rise to unique collider signatures that distinguish them from standard collider searches. In conclusion, this framework unifies the origin of proton decay and DM stability through the breaking of \UBL symmetry to \Z4, while also predicting distinctive collider signals that set it apart from conventional scenarios.

       
\section{Acknowledgments}
 \noindent We would like to thank Martin Hirsch, Oleg Popov, and Duttatreya Malayaja for useful discussions. Also, we acknowledge Salvador Centelles Chuliá for his contributions during the preliminary stage of this work. Authors would like to acknowledge the SARAH~\cite{Staub:2015kfa}, SPheno~\cite{Porod:2011nf}, micrOMEGAs~\cite{Belanger:2014vza}, and MadGraph5~\cite{Alwall:2014hca} packages, which have been used to perform the numerical analysis. RK acknowledges the funding support from the CSIR SRF-NET fellowship. 
\appendix  

\section{Fermionic DM Scenario} \label{sec:fer_dm}
In this section, we briefly discuss an alternative UV-completion of the dim-6 effective operator $[d u][ue]$, featuring a fermionic DM candidate. This scenario also leads to the two-body proton decay process $p \rightarrow e^+ \pi^0$. The corresponding Feynman diagram illustrating this decay is shown in Fig.~\ref{fig:pdfeyn_fermi}.
\begin{figure}[!h]
    \centering
    \begin{tikzpicture} 
\fill[black!100] (-3.2,0) ellipse (0.4 and 2.4);
\fill[black!100] (2.4,-1.8) ellipse (0.25 and 0.9);
\begin{feynman}
\vertex (cntf);
\vertex [above = 1cm of cntf] (cnt);
\vertex [above = 1.5cm of cnt] (sigv);
\vertex [below = 1.5cm of cnt] (chiv);
\vertex [above = 2.5cm of cnt] (sig);
\vertex [below = 2.5cm of cnt] (chi);
\vertex [above left = 1.5cm of cnt] (N1);
\vertex [below left = 1.5cm of cnt] (N31);
\vertex [above right = 1.5cm of cnt] (N2);
\vertex [below right = 1.5cm of cnt] (N32);
\vertex [below = 3cm of cnt] (fu);
\vertex [left=2cm of N1] (d);
\vertex [left=2cm of N31] (u2);
\vertex [left=3cm of fu] (u1);
\vertex [left=3.6cm of cntf] (p) {$p$};
\vertex [right=1.2cm of N2] (e) {$e^+$};
\vertex [right=1.2cm of N32] (u2rf);
\vertex [below = 1.2cm of u2rf] (u2r);
\vertex [below=1.0cm of u2r] (u1r);
\vertex [right = 2.6cm of chi] (pi) {$\pi^0$};

\diagram* {
(u1) -- [ fermion, edge label = $u$] (fu) -- [ fermion, edge label = $u$] (u1r),
(d) -- [ fermion, edge label = $d$] (N1),
(sigv) -- [anti fermion, edge label = $N_2$] (N2),
(sigv) -- [fermion, edge label' = $N_1$] (N1),
(e) -- [fermion] (N2),
(u2) -- [ fermion, edge label = $u$] (N31),
(N32) -- [anti fermion, edge label = $u$] (u2r),
(N31) -- [anti fermion, edge label' = $D_1$] (chiv) -- [anti fermion, edge label' = $D_2$] (N32),
(N1) -- [anti charged scalar, edge label' = $S_1$] (N31),
(N2) -- [ charged scalar, edge label = $\phi^-$] (N32),
(sig) -- [ anti charged scalar, insertion = 0.01, edge label' = $\sigma$] (sigv),
(chi) -- [ anti charged scalar, insertion = 0.01, edge label' = $\chi$] (chiv),
};
\end{feynman}
\end{tikzpicture}
\caption{One-loop Feynman diagram for proton decay via new dark sector particles in the case of fermionic DM.}
\label{fig:pdfeyn_fermi}
\end{figure}
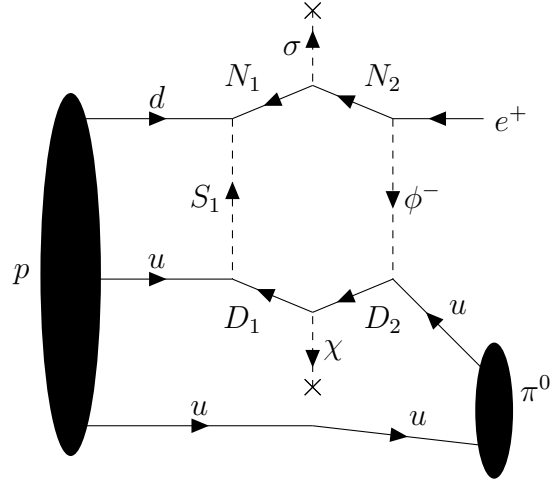
    \begin{table}[!h]
\begin{center}
\begin{tabular}{| c || c || c || c|}
  \hline
& Fields  & $SU(3)_C \otimes SU(2)_L \otimes U(1)_Y$           & $U(1)_{B+L}$   $\to$ {\red  $Z_{4}$}  \       \\
\hline \hline
\ \multirow{6}{*}{\rotatebox{90}{SM}} \ & $Q_i$ \ \   & $(3, 2, \frac{1}{6})$ \ \   &  $\frac{1}{3}$ $\to$ {\red    $\omega^{2}$ }\\
 & $L_i$ \ \   & $(1, 2, -\frac{1}{2})$\ \  &  $1$  $\to$ {\red    $\omega^{2}$ } \\
& $u_i$ \ \   & $(3, 1, \frac{2}{3})$  \ \  &  $\frac{1}{3}$ $\to$ {\red     $\omega^{2}$} \\ 
& $e_i$  \ \  & $(1, 1, -1)$  \ \   &  $1$  $\to$ {\red     $\omega^{2}$}      \\
& $d_i$ \ \   & $(3, 1, -\frac{1}{3})$ \ \  &  $\frac{1}{3}$  $\to$ {\red    $\omega^{2}$} \\
& $H$\ \    & $(1, 2, \frac{1}{2})$  \ \   &  $0$ $\to$ {\red     $1$} \\
\hline \hline
 \ \multirow{6}{*}{\rotatebox{90}{BSM }} \  
& $(\boldsymbol{N_{1},N_{2}})$ \ \   & $(\boldsymbol{1, 1, 0})$ \ \  &  $(\boldsymbol{-\frac{1}{2},\frac{5}{6}})$  $\to$ {\red    $(\boldsymbol{\omega,\omega})$} \\
& $(D_1,D_2)$ \ \   & $(3, 1, -\frac{1}{3})$ \ \  &  $(-\frac{1}{6},\frac{1}{2})$  $\to$ {\red     $(\omega^3,\omega^3)$} \\
& $S_1$ \ \   & $(\bar{3}, 1, \frac{1}{3})$ \ \  &  $\frac{1}{6}$  $\to$ {\red   $\omega$}  \\
& $\sigma$ \ \   & $(1, 1, 0)$ \ \  &  $\frac{4}{3}$  $\to$ {\red     $1$ } \\
& $\chi$ \ \   & $(1, 1, 0)$ \ \  &  $\frac{2}{3}$  $\to$ {\red     $1$} \\
& $\phi^-$ \ \   & $(1, 1, -1)$ \ \  &  $\frac{1}{6}$  $\to$ {\red     $\omega$ } \\        
\hline 
  \end{tabular}
\end{center}
\caption{Particle content and their transformation under different symmetries in the case of fermionic DM.}
  \label{tab:ferm}
\end{table}
 The charge assignments of the particles are chosen such that the color neutral fermions $(N_1,N_2)$ are also electromagnetically neutral and can be a viable DM candidate. In this new charge assignment, we have vector-like heavy quark states $(D_1, D_2)$ and scalar leptoquark $S_1$ as well as charged scalar $\phi^-$. The charges of SM particles, along with the scalars $\chi$ and $\sigma$, under SM symmetries and \UBL, remain unchanged. However, BSM particles running inside the loop have different charge assignments.
 Similar to the scalar case, the particles mediating loop-induced proton decay are odd under the residual \Z4 symmetry and belong to the dark sector, as summarized in Tab.~\ref{tab:ferm}. The proton lifetime is primarily determined by the masses of these dark sector particles and their Yukawa couplings to SM particles, leading to proton decay constraints that closely resemble the scalar DM case. However, the DM and collider phenomenology exhibit significant differences. Unlike scalar DM, fermionic DM has distinct relic density production channels and DD prospects, leading to different constraints on the viable parameter space. Furthermore, the UV completion involves new BSM particles, including a scalar leptoquark $S_1$, a charged scalar $\phi^-$, and heavy vector-like quarks $(D_1, D_2)$. These particles not only modify the collider signatures compared to the scalar DM case but also introduce new experimental constraints on their masses. This alternative UV framework thus provides distinct signatures which can be used to distinguish it from the scalar DM case discussed in the main text.
 

\section{Computation of Proton Decay Rate}
\label{app:proton}
\begin{figure}[!h]
    \centering

\begin{tikzpicture}
\begin{feynman}
\vertex (c);
\vertex [above = 3 cm of c] (chi);
\vertex [above = 2 cm of c] (chi0);
\vertex [above = 0.1 cm of chi0] (chi0p){\(E\)};
\vertex [above right = 2 cm of c] (u20);
\vertex [above right = 3 cm of c] (u2){\(e\)};
\vertex [above left = 2 cm of c] (u10);
\vertex [above left = 3 cm of c] (u1){\(d\)};
\vertex [below right = 2 cm of c] (e0);
\vertex [below right = 3 cm of c] (e){\(u\)};
\vertex [below left = 2 cm of c] (d0);
\vertex [below left = 3 cm of c] (d){\(u\)};
\vertex [below = 2 cm of c] (sigma0);
\vertex [below = 0.12 cm of sigma0] (sigma0p){\(U\)};
\vertex [below = 3 cm of c] (sigma);
\diagram* {
(u10) -- [anti fermion,  momentum'={[arrow shorten=0.28mm,arrow distance=2.25mm, red]\(p_2\)}] (u1),
(u20) -- [anti fermion, rmomentum={[arrow shorten=0.28mm,arrow distance=2.25mm, red]\(p_3\)}] (u2),
(d0) -- [anti fermion,  momentum={[arrow shorten=0.28mm,arrow distance=2.25mm, red]\(p_1\)}] (d),
(e0) -- [anti fermion,  rmomentum'={[arrow shorten=0.28mm,arrow distance=2.mm, red]\(p_4\)}] (e),

(u10) -- [anti fermion] (chi0),
(u20) -- [fermion,  rmomentum'={[arrow shorten=0.28mm,arrow distance=2.mm, red]\(k-p_{43}\)}] (chi0),
(d0) -- [anti fermion] (sigma0),
(e0) -- [fermion, momentum={[arrow shorten=0.28mm,arrow distance=2.25mm, red]\(k\)}] (sigma0),

(u10) -- [ anti charged scalar, edge label = $\tilde{S}_1$, rmomentum={[arrow shorten=0.23mm,arrow distance=2.25mm, red]\(k-p_1\)}] (d0),
(u20) -- [ charged scalar, edge label' = $\zeta$, momentum'={[arrow shorten=0.23mm,arrow distance=2.25mm, red]\(k-p_4\)}] (e0),

};
\end{feynman}
\end{tikzpicture}
\caption{One-loop calculation of proton decay, where $p_i$ are external momenta and $k$ is the internal loop momentum.}
    \label{fig:loopcal}
\end{figure}
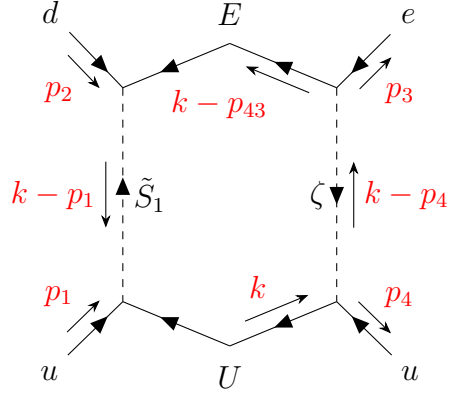

Here, we focus on the calculation of the proton decay rate, which is mediated by a one-loop box diagram. The effective operator in our model is $[d u][ue]$, which encapsulates the baryon and lepton number violating interactions responsible for proton decay. The UV-completion of this operator is shown in Fig.~\ref{fig:loopcal}, where the dynamics of the loop level process are depicted. 
In the box diagram shown in Fig.~\ref{fig:loopcal}, external momenta are labeled as $p_i$ ($i=1,2,3,4$) and the internal momentum as $k$. The loop factor for the proton decay can be computed as
    \begin{align} \label{eq:cexp}
        \mathcal{C}=-|Y|^4 M_E M_U \mathcal{I} ,
    \end{align}
 where $\mathcal{I}$ is the loop integral given by
 \begin{align} \label{eq:loopfac}
     \mathcal{I}=\int_k \frac{d^4k }{(2\pi)^4 i} \frac{1}{k^2-M_U^2}\frac{1}{(k-p_4)^2-m_{\zeta}^2}\frac{1}{(k-p_{43})^2-M_E^2}\frac{1}{(k-p_1)^2-M_{\tilde{S}_1}^2},
 \end{align}
  and $|Y|\equiv (Y_1 Y_2 Y_3 Y_4)^{1/4}$, $p_{43}\equiv p_3+p_4$.
Using the loop factor given in Eq.~\eqref{eq:cexp}, the proton decay rate for the process, $p \rightarrow e^+ \pi^0$ can be computed as follows
\begin{equation}
\Gamma\left(p \rightarrow  e^{+} \pi^0  \right)=\frac{m_p}{32 \pi}\left[1-\frac{m_{\pi^{0}}^2}{m_p^2}\right]^2\left|W_0 \mathcal{C}\right|^2,
\end{equation}
where $W_0$ is the hadronic matrix element, $W_0 \equiv -0.131$ $\rm{GeV}^2$~\cite{Aoki:2017puj}.

\vspace{-0.75cm}

\section{Anomalies of the \UBL Symmetry} \label{sec:anomaly}
In this section, we briefly discuss the anomalies induced by the fermions of our model under the SM gauge symmetries and the \UBL symmetry. As \UBL is not gauged in our framework, we only list the relevant anomaly cancellation conditions for completeness. We begin by listing the anomalies associated with the SM gauge symmetry.
\begin{align}
    \left[ SU(3)_C \right]^2 \left[U(1)_Y \right] &= 3 \sum_i Y_{Q'_i}  - 3 \sum_j Y_{q'_j} , \\
    \left[ SU(2)_L \right]^2 \left[U(1)_Y \right] &=  \sum_i Y_{L_i} + 3 \sum_j Y_{Q_j}, \\
    \left[ U(1)_Y \right]^3  &=  \sum_{i,j}\left( Y^3_{L'_i} + 3Y^3_{Q'_j} \right) -  \sum_{i,j}\left( Y^3_{e'_i} + 3Y^3_{q'_j} \right), \\ 
    \left[ G \right]^2 \left[U(1)_Y \right] &=  \sum_{i,j}\left( Y_{L'_i} + 3Y_{Q'_j} \right) -  \sum_{i,j}\left( Y_{e'_i} + 3Y_{q'_j} \right). 
    \end{align}
     where $Y$ is the hypercharge of the particles. We denote fermions as, $Q'_i \equiv (Q_i, U_L)$,
     $q'_j \equiv (q_j, U_R)$, $L'_i \equiv (L_i, E_L)$, $e'_i \equiv (e_i, E_R)$, and $i,j$ denote the generation indices of SM fermions. The SM fermions $Q_i, L_j$ are doublets under $SU(2)_L$, while the SM fermions $q_i,e_j$ and the BSM fermions
     $U_L,U_R,E_L,E_R$ are singlets of $SU(2)_L$.
     The factor $3$ accounts for the color multiplicity.    
     Next, we present the additional anomalies induced by the \UBL symmetry.
    \begin{align}
    \left[ SU(3)_C \right]^2 \left[U(1)_{B+L} \right] &= 3 \sum_i X_{Q'_i}  - 3 \sum_j X_{q'_j}, \\
    \left[ SU(2)_L \right]^2 \left[U(1)_{B+L} \right] &=  \sum_i X_{L_i} + 3 \sum_j X_{Q_j}, \\
    \left[ U(1)_Y \right]^2 \left[ U(1)_{B+L} \right]  &=  \sum_{i,j}\left( Y^2_{L'_i} X_{L'_i} + 3 Y^2_{Q'_j} X_{Q'_j} \right) -  \sum_{i,j}\left(Y^2_{e'_i} X_{e'_i} + 3 Y^2_{q'_j} X_{q'_j} \right), \\ 
    \left[ U(1)_Y \right] \left[ U(1)_{B+L} \right]^2  &=  \sum_{i,j}\left( Y_{L'_i} X^2_{L'_i} + 3 Y_{Q'_j} X^2_{Q'_j} \right) -  \sum_{i,j}\left(Y_{e'_i} X^2_{e'_i} + 3 Y_{q'_j} X^2_{q'_j} \right), \\ 
    \left[ U(1)_{B+L} \right]^3  &=  \sum_{i,j}\left( X^3_{L'_i} + 3X^3_{Q'_j} \right) -  \sum_{i,j}\left( X^3_{e'_i} + 3X^3_{q'_j} \right), \\ 
    \left[ G \right]^2 \left[U(1)_{B+L} \right] &=  \sum_{i,j}\left( X_{L'_i} + 3X_{Q'_j} \right) -  \sum_{i,j}\left( X_{e'_i} + 3X_{q'_j} \right). 
\end{align}
Here, $X$ denotes the $B+L$ charge of the particles.
 See Tab.~\ref{tab:particle} for the charges of particles in our model under SM gauge symmetries and  \UBL global symmetry.

Since, in our model, the extra fermions $(U_L, U_R)$ and $(E_L, E_R)$ are vector-like under the SM gauge symmetry, their contributions to the anomaly conditions in Eqs. ({\color{red}C1–C4}) vanish identically. In addition, the SM fermions are vector-like with respect to the newly introduced \UBL symmetry, and hence their contributions to the anomaly conditions in Eqs. ({\color{red}C5}), ({\color{red}C9}), and ({\color{red}C10}) goes to zero.
As a result, the remaining non-vanishing anomaly contributions are given by
\begin{align}
     \left[ SU(3)_C \right]^2 \left[U(1)_{B+L} \right] &= 3 X_{U_L} -3 X_{U_R}, \\
    \left[ SU(2)_L \right]^2 \left[U(1)_{B+L} \right] &=  \sum_i X_{L_i} + 3 \sum_j X_{Q_j}, \\
     \left[ U(1)_Y \right]^2 \left[ U(1)_{B+L} \right] & =  \sum_{i,j}\left( Y^2_{L'_i} X_{L'_i} + 3 Y^2_{Q'_j} X_{Q'_j} \right) -  \sum_{i,j}\left(Y^2_{e'_i} X_{e'_i} + 3 Y^2_{q'_j} X_{q'_j} \right), \\ 
    \left[ U(1)_Y \right] \left[ U(1)_{B+L} \right]^2 & =  \sum_{i,j}\left( Y_{L'_i} X^2_{L'_i} + 3 Y_{Q'_j} X^2_{Q'_j} \right) -  \sum_{i,j}\left(Y_{e'_i} X^2_{e'_i} + 3 Y_{q'_j} X^2_{q'_j} \right), \\ 
    \left[ U(1)_{B+L} \right]^3  &=   X^3_{E_L} + 3 X^3_{U_L}- X^3_{E_R}- 3X^3_{U_R}, \\ 
    \left[ G \right]^2 \left[U(1)_{B+L} \right] &=  X_{E_L} + 3 X_{U_L}- X_{E_R}- 3X_{U_R}. 
\end{align}
The above mentioned anomaly equations are non-vanishing. Therefore, in order to consistently gauge the \UBL symmetry, one must introduce additional BSM fermions with appropriate charge assignments such that the individual anomaly contributions in Eqs.~({\color{red}C11–C16}) cancel separately. In the present work, we treat \UBL as a global symmetry and do not delve into the details of anomaly free realizations required for a gauged \UBL scenario.
\FloatBarrier
\bibliographystyle{utphys}
\bibliography{references} 

\providecommand{\href}[2]{#2}\begingroup\raggedright\begin{thebibliography}{10}

\bibitem{Weinberg:1979sa}
S.~Weinberg, ``{Baryon and Lepton Nonconserving Processes},''
  \href{http://dx.doi.org/10.1103/PhysRevLett.43.1566}{{\em Phys. Rev. Lett.}
  {\bfseries 43} (1979) 1566--1570}.

\bibitem{Wilczek:1979hc}
F.~Wilczek and A.~Zee, ``{Operator Analysis of Nucleon Decay},''
  \href{http://dx.doi.org/10.1103/PhysRevLett.43.1571}{{\em Phys. Rev. Lett.}
  {\bfseries 43} (1979) 1571--1573}.

\bibitem{Abbott:1980zj}
L.~F. Abbott and M.~B. Wise, ``{The Effective Hamiltonian for Nucleon Decay},''
  \href{http://dx.doi.org/10.1103/PhysRevD.22.2208}{{\em Phys. Rev. D}
  {\bfseries 22} (1980) 2208}.

\bibitem{Pati:1973uk}
J.~C. Pati and A.~Salam, ``{Unified Lepton-Hadron Symmetry and a Gauge Theory
  of the Basic Interactions},''
  \href{http://dx.doi.org/10.1103/PhysRevD.8.1240}{{\em Phys. Rev. D}
  {\bfseries 8} (1973) 1240--1251}.

\bibitem{Georgi:1974sy}
H.~Georgi and S.~L. Glashow, ``{Unity of All Elementary Particle Forces},''
  \href{http://dx.doi.org/10.1103/PhysRevLett.32.438}{{\em Phys. Rev. Lett.}
  {\bfseries 32} (1974) 438--441}.

\bibitem{Fritzsch:1974nn}
H.~Fritzsch and P.~Minkowski, ``{Unified Interactions of Leptons and
  Hadrons},'' \href{http://dx.doi.org/10.1016/0003-4916(75)90211-0}{{\em Annals
  Phys.} {\bfseries 93} (1975) 193--266}.

\bibitem{Super-Kamiokande:2020wjk}
{\bfseries Super-Kamiokande} Collaboration, A.~Takenaka {\em et~al.}, ``{Search
  for proton decay via $p\to e^+\pi^0$ and $p\to \mu^+\pi^0$ with an enlarged
  fiducial volume in Super-Kamiokande I-IV},''
  \href{http://dx.doi.org/10.1103/PhysRevD.102.112011}{{\em Phys. Rev. D}
  {\bfseries 102} no.~11, (2020) 112011},
  \href{http://arxiv.org/abs/2010.16098}{{\ttfamily arXiv:2010.16098
  [hep-ex]}}.

\bibitem{Super-Kamiokande:2022egr}
{\bfseries Super-Kamiokande} Collaboration, R.~Matsumoto {\em et~al.},
  ``{Search for proton decay via $p\rightarrow \mu^+K^0$ in 0.37 megaton-years
  exposure of Super-Kamiokande},''
  \href{http://dx.doi.org/10.1103/PhysRevD.106.072003}{{\em Phys. Rev. D}
  {\bfseries 106} no.~7, (2022) 072003},
  \href{http://arxiv.org/abs/2208.13188}{{\ttfamily arXiv:2208.13188
  [hep-ex]}}.

\bibitem{DUNE:2020ypp}
{\bfseries DUNE} Collaboration, B.~Abi {\em et~al.}, ``{Deep Underground
  Neutrino Experiment (DUNE), Far Detector Technical Design Report, Volume II:
  DUNE Physics},'' \href{http://arxiv.org/abs/2002.03005}{{\ttfamily
  arXiv:2002.03005 [hep-ex]}}.

\bibitem{Hyper-Kamiokande:2018ofw}
{\bfseries Hyper-Kamiokande} Collaboration, K.~Abe {\em et~al.},
  ``{Hyper-Kamiokande Design Report},''
  \href{http://arxiv.org/abs/1805.04163}{{\ttfamily arXiv:1805.04163
  [physics.ins-det]}}.

\bibitem{JUNO:2021vlw}
{\bfseries JUNO} Collaboration, A.~Abusleme {\em et~al.}, ``{JUNO physics and
  detector},'' \href{http://dx.doi.org/10.1016/j.ppnp.2021.103927}{{\em Prog.
  Part. Nucl. Phys.} {\bfseries 123} (2022) 103927},
  \href{http://arxiv.org/abs/2104.02565}{{\ttfamily arXiv:2104.02565
  [hep-ex]}}.

\bibitem{Helo:2019yqp}
J.~C. Helo, M.~Hirsch, and T.~Ota, ``{Proton decay at one loop},''
  \href{http://dx.doi.org/10.1103/PhysRevD.99.095021}{{\em Phys. Rev. D}
  {\bfseries 99} no.~9, (2019) 095021},
  \href{http://arxiv.org/abs/1904.00036}{{\ttfamily arXiv:1904.00036
  [hep-ph]}}.

\bibitem{Dorsner:2022twk}
I.~Dor\v{s}ner, S.~Fajfer, and O.~Sumensari, ``{Triple-leptoquark interactions
  for tree- and loop-level proton decays},''
  \href{http://dx.doi.org/10.1007/JHEP05(2022)183}{{\em JHEP} {\bfseries 05}
  (2022) 183}, \href{http://arxiv.org/abs/2202.08287}{{\ttfamily
  arXiv:2202.08287 [hep-ph]}}.

\bibitem{Nomura:2024zca}
T.~Nomura and O.~Popov, ``{Extended scotogenic model of neutrino mass and
  proton decay},'' \href{http://dx.doi.org/10.1103/PhysRevD.110.075035}{{\em
  Phys. Rev. D} {\bfseries 110} no.~7, (2024) 075035},
  \href{http://arxiv.org/abs/2406.00651}{{\ttfamily arXiv:2406.00651
  [hep-ph]}}.

\bibitem{Kang:2024oyf}
S.~K. Kang and O.~Popov, ``{Pathways to proton's stability via naturally small
  neutrino masses},'' \href{http://arxiv.org/abs/2412.20723}{{\ttfamily
  arXiv:2412.20723 [hep-ph]}}.

\bibitem{ODonnell:1993kdg}
P.~J. O'Donnell and U.~Sarkar, ``{Three lepton decay mode of the proton},''
  \href{http://dx.doi.org/10.1016/0370-2693(93)90667-7}{{\em Phys. Lett. B}
  {\bfseries 316} (1993) 121--126},
  \href{http://arxiv.org/abs/hep-ph/9307254}{{\ttfamily arXiv:hep-ph/9307254}}.

\bibitem{Babu:2012iv}
K.~S. Babu and R.~N. Mohapatra, ``{B-L Violating Proton Decay Modes and New
  Baryogenesis Scenario in SO(10)},''
  \href{http://dx.doi.org/10.1103/PhysRevLett.109.091803}{{\em Phys. Rev.
  Lett.} {\bfseries 109} (2012) 091803},
  \href{http://arxiv.org/abs/1207.5771}{{\ttfamily arXiv:1207.5771 [hep-ph]}}.

\bibitem{Lehman:2014jma}
L.~Lehman, ``{Extending the Standard Model Effective Field Theory with the
  Complete Set of Dimension-7 Operators},''
  \href{http://dx.doi.org/10.1103/PhysRevD.90.125023}{{\em Phys. Rev. D}
  {\bfseries 90} no.~12, (2014) 125023},
  \href{http://arxiv.org/abs/1410.4193}{{\ttfamily arXiv:1410.4193 [hep-ph]}}.

\bibitem{Bhattacharya:2015vja}
S.~Bhattacharya and J.~Wudka, ``{Dimension-seven operators in the standard
  model with right handed neutrinos},''
  \href{http://dx.doi.org/10.1103/PhysRevD.94.055022}{{\em Phys. Rev. D}
  {\bfseries 94} no.~5, (2016) 055022},
  \href{http://arxiv.org/abs/1505.05264}{{\ttfamily arXiv:1505.05264
  [hep-ph]}}. [Erratum: Phys.Rev.D 95, 039904 (2017)].

\bibitem{Liao:2016hru}
Y.~Liao and X.-D. Ma, ``{Renormalization Group Evolution of Dimension-seven
  Baryon- and Lepton-number-violating Operators},''
  \href{http://dx.doi.org/10.1007/JHEP11(2016)043}{{\em JHEP} {\bfseries 11}
  (2016) 043}, \href{http://arxiv.org/abs/1607.07309}{{\ttfamily
  arXiv:1607.07309 [hep-ph]}}.

\bibitem{Liao:2016qyd}
Y.~Liao and X.-D. Ma, ``{Operators up to Dimension Seven in Standard Model
  Effective Field Theory Extended with Sterile Neutrinos},''
  \href{http://dx.doi.org/10.1103/PhysRevD.96.015012}{{\em Phys. Rev. D}
  {\bfseries 96} no.~1, (2017) 015012},
  \href{http://arxiv.org/abs/1612.04527}{{\ttfamily arXiv:1612.04527
  [hep-ph]}}.

\bibitem{Hambye:2017qix}
T.~Hambye and J.~Heeck, ``{Proton decay into charged leptons},''
  \href{http://dx.doi.org/10.1103/PhysRevLett.120.171801}{{\em Phys. Rev.
  Lett.} {\bfseries 120} no.~17, (2018) 171801},
  \href{http://arxiv.org/abs/1712.04871}{{\ttfamily arXiv:1712.04871
  [hep-ph]}}.

\bibitem{Fonseca:2018ehk}
R.~M. Fonseca, M.~Hirsch, and R.~Srivastava, ``{$\Delta L = 3$ processes:
  Proton decay and the LHC},''
  \href{http://dx.doi.org/10.1103/PhysRevD.97.075026}{{\em Phys. Rev. D}
  {\bfseries 97} no.~7, (2018) 075026},
  \href{http://arxiv.org/abs/1802.04814}{{\ttfamily arXiv:1802.04814
  [hep-ph]}}.

\bibitem{Beneito:2023xbk}
A.~B. Beneito, I, J.~Gargalionis, J.~Herrero-Garcia, A.~Santamaria, and M.~A.
  Schmidt, ``{An EFT approach to baryon number violation: lower limits on the
  new physics scale and correlations between nucleon decay modes},''
  \href{http://dx.doi.org/10.1007/JHEP07(2024)004}{{\em JHEP} {\bfseries 07}
  (2024) 004}, \href{http://arxiv.org/abs/2312.13361}{{\ttfamily
  arXiv:2312.13361 [hep-ph]}}.

\bibitem{IBeneito:2025nby}
A.~B. I~Beneito, J.~Gargalionis, J.~Herrero-Garcia, and M.~A. Schmidt,
  ``{Squeezing Proton Decay and Neutrino Masses: Upper Bounds on Standard Model
  Extensions},'' \href{http://arxiv.org/abs/2503.20928}{{\ttfamily
  arXiv:2503.20928 [hep-ph]}}.

\bibitem{Liao:2025vlj}
Y.~Liao, X.-D. Ma, and H.-L. Wang, ``{New chiral structures for nucleon baryon
  number violating decays},'' \href{http://arxiv.org/abs/2504.14855}{{\ttfamily
  arXiv:2504.14855 [hep-ph]}}.

\bibitem{Planck:2018vyg}
{\bfseries Planck} Collaboration, N.~Aghanim {\em et~al.}, ``{Planck 2018
  results. VI. Cosmological parameters},''
  \href{http://dx.doi.org/10.1051/0004-6361/201833910}{{\em Astron. Astrophys.}
  {\bfseries 641} (2020) A6}, \href{http://arxiv.org/abs/1807.06209}{{\ttfamily
  arXiv:1807.06209 [astro-ph.CO]}}. [Erratum: Astron.Astrophys. 652, C4
  (2021)].

\bibitem{Ma:2006km}
E.~Ma, ``{Verifiable radiative seesaw mechanism of neutrino mass and dark
  matter},'' \href{http://dx.doi.org/10.1103/PhysRevD.73.077301}{{\em Phys.
  Rev. D} {\bfseries 73} (2006) 077301},
  \href{http://arxiv.org/abs/hep-ph/0601225}{{\ttfamily arXiv:hep-ph/0601225}}.

\bibitem{Kang:2019sab}
S.~K. Kang, O.~Popov, R.~Srivastava, J.~W.~F. Valle, and C.~A. Vaquera-Araujo,
  ``{Scotogenic dark matter stability from gauged matter parity},''
  \href{http://dx.doi.org/10.1016/j.physletb.2019.135013}{{\em Phys. Lett. B}
  {\bfseries 798} (2019) 135013},
  \href{http://arxiv.org/abs/1902.05966}{{\ttfamily arXiv:1902.05966
  [hep-ph]}}.

\bibitem{Leite:2020wjl}
J.~Leite, A.~Morales, J.~W.~F. Valle, and C.~A. Vaquera-Araujo, ``{Scotogenic
  dark matter and Dirac neutrinos from unbroken gauged B-L symmetry},''
  \href{http://dx.doi.org/10.1016/j.physletb.2020.135537}{{\em Phys. Lett. B}
  {\bfseries 807} (2020) 135537},
  \href{http://arxiv.org/abs/2003.02950}{{\ttfamily arXiv:2003.02950
  [hep-ph]}}.

\bibitem{CentellesChulia:2022vpz}
S.~Centelles~Chuli\'a, R.~Srivastava, and S.~Yadav, ``{CDF-II W boson mass in
  the Dirac Scotogenic model},''
  \href{http://dx.doi.org/10.1142/S0217732323500499}{{\em Mod. Phys. Lett. A}
  {\bfseries 38} no.~7, (2023) },
  \href{http://arxiv.org/abs/2206.11903}{{\ttfamily arXiv:2206.11903
  [hep-ph]}}.

\bibitem{Batra:2022pej}
A.~Batra, S.~K. A, S.~Mandal, H.~Prajapati, and R.~Srivastava, ``{CDF-II
  W-boson mass anomaly in the canonical Scotogenic neutrino\textendash{}dark
  matter model},'' \href{http://dx.doi.org/10.1142/S0217732323500906}{{\em Mod.
  Phys. Lett. A} {\bfseries 38} no.~18n19, (2023) 2350090},
  \href{http://arxiv.org/abs/2204.11945}{{\ttfamily arXiv:2204.11945
  [hep-ph]}}.

\bibitem{Kumar:2023moh}
R.~Kumar, P.~Mishra, M.~K. Behera, R.~Mohanta, and R.~Srivastava,
  ``{Predictions from scoto-seesaw with A4 modular symmetry},''
  \href{http://dx.doi.org/10.1016/j.physletb.2024.138635}{{\em Phys. Lett. B}
  {\bfseries 853} (2024) 138635},
  \href{http://arxiv.org/abs/2310.02363}{{\ttfamily arXiv:2310.02363
  [hep-ph]}}.

\bibitem{Kumar:2024jot}
R.~Kumar, N.~Nath, and R.~Srivastava, ``{Cutting the~Scotogenic Loop: $A_4$
  Flavor Symmetry to~$Z_2$ Dark Symmetry},''
  \href{http://dx.doi.org/10.1007/978-981-97-0289-3_331}{{\em Springer Proc.
  Phys.} {\bfseries 304} (2024) 1183--1185}.

\bibitem{Kumar:2024zfb}
R.~Kumar, N.~Nath, and R.~Srivastava, ``{Cutting the scotogenic loop: adding
  flavor to dark matter},''
  \href{http://dx.doi.org/10.1007/JHEP12(2024)036}{{\em JHEP} {\bfseries 12}
  (2024) 036}, \href{http://arxiv.org/abs/2406.00188}{{\ttfamily
  arXiv:2406.00188 [hep-ph]}}.

\bibitem{CentellesChulia:2024iom}
S.~Centelles~Chuli\'a, R.~Srivastava, and S.~Yadav, ``{Comprehensive
  Phenomenology of the Dirac Scotogenic Model: Novel Low Mass Dark Matter},''
  \href{http://arxiv.org/abs/2409.18513}{{\ttfamily arXiv:2409.18513
  [hep-ph]}}.

\bibitem{Garnica:2024wur}
Y.~Garnica, A.~Morales, and C.~A. Vaquera-Araujo, ``{Scotogenic dark matter
  from gauged $B-L$},'' \href{http://arxiv.org/abs/2411.13756}{{\ttfamily
  arXiv:2411.13756 [hep-ph]}}.

\bibitem{Singh:2025jtn}
L.~Singh, R.~Srivastava, S.~Verma, and S.~Yadav, ``{Type-III Scotogenic Model:
  Inflation, Dark Matter and Collider Phenomenology},''
  \href{http://arxiv.org/abs/2501.13171}{{\ttfamily arXiv:2501.13171
  [hep-ph]}}.

\bibitem{Lozano:2025tst}
V.~M. Lozano, G.~Sanchez~Garcia, and J.~W.~F. Valle, ``{Collider signatures of
  fermionic scotogenic dark matter},''
  \href{http://arxiv.org/abs/2502.05270}{{\ttfamily arXiv:2502.05270
  [hep-ph]}}.

\bibitem{Kumar:2025cte}
R.~Kumar, N.~Nath, R.~Srivastava, and S.~Yadav, ``{Dirac Scoto Inverse-Seesaw
  from $A_4$ Flavor Symmetry},''
  \href{http://arxiv.org/abs/2505.01407}{{\ttfamily arXiv:2505.01407
  [hep-ph]}}.

\bibitem{Reig:2018yfd}
M.~Reig and R.~Srivastava, ``{Spontaneous proton decay and the origin of
  Peccei\textendash{}Quinn symmetry},''
  \href{http://dx.doi.org/10.1016/j.physletb.2019.01.008}{{\em Phys. Lett. B}
  {\bfseries 790} (2019) 134--139},
  \href{http://arxiv.org/abs/1809.02093}{{\ttfamily arXiv:1809.02093
  [hep-ph]}}.

\bibitem{CMS:2012qbp}
{\bfseries CMS} Collaboration, S.~Chatrchyan {\em et~al.}, ``{Observation of a
  New Boson at a Mass of 125 GeV with the CMS Experiment at the LHC},''
  \href{http://dx.doi.org/10.1016/j.physletb.2012.08.021}{{\em Phys. Lett. B}
  {\bfseries 716} (2012) 30--61},
  \href{http://arxiv.org/abs/1207.7235}{{\ttfamily arXiv:1207.7235 [hep-ex]}}.

\bibitem{ATLAS:2012yve}
{\bfseries ATLAS} Collaboration, G.~Aad {\em et~al.}, ``{Observation of a new
  particle in the search for the Standard Model Higgs boson with the ATLAS
  detector at the LHC},''
  \href{http://dx.doi.org/10.1016/j.physletb.2012.08.020}{{\em Phys. Lett. B}
  {\bfseries 716} (2012) 1--29},
  \href{http://arxiv.org/abs/1207.7214}{{\ttfamily arXiv:1207.7214 [hep-ex]}}.

\bibitem{ParticleDataGroup:2024cfk}
{\bfseries Particle Data Group} Collaboration, S.~Navas {\em et~al.}, ``{Review
  of particle physics},''
  \href{http://dx.doi.org/10.1103/PhysRevD.110.030001}{{\em Phys. Rev. D}
  {\bfseries 110} no.~3, (2024) 030001}.

\bibitem{Cline:2013gha}
J.~M. Cline, K.~Kainulainen, P.~Scott, and C.~Weniger, ``{Update on scalar
  singlet dark matter},''
  \href{http://dx.doi.org/10.1103/PhysRevD.88.055025}{{\em Phys. Rev. D}
  {\bfseries 88} (2013) 055025},
  \href{http://arxiv.org/abs/1306.4710}{{\ttfamily arXiv:1306.4710 [hep-ph]}}.
  [Erratum: Phys.Rev.D 92, 039906 (2015)].

\bibitem{Aoki:2017puj}
Y.~Aoki, T.~Izubuchi, E.~Shintani, and A.~Soni, ``{Improved lattice computation
  of proton decay matrix elements},''
  \href{http://dx.doi.org/10.1103/PhysRevD.96.014506}{{\em Phys. Rev. D}
  {\bfseries 96} no.~1, (2017) 014506},
  \href{http://arxiv.org/abs/1705.01338}{{\ttfamily arXiv:1705.01338
  [hep-lat]}}.

\bibitem{Bharadwaj:2024crt}
P.~Bharadwaj, R.~Kumar, H.~K. Prajapati, R.~Srivastava, and S.~Yadav, ``{Dark
  Matter Escaping Direct Detection Runs into Higgs Mass Hierarchy Problem},''
  \href{http://arxiv.org/abs/2412.13301}{{\ttfamily arXiv:2412.13301
  [hep-ph]}}.

\bibitem{LZ:2022lsv}
{\bfseries LZ} Collaboration, J.~Aalbers {\em et~al.}, ``{First Dark Matter
  Search Results from the LUX-ZEPLIN (LZ) Experiment},''
  \href{http://dx.doi.org/10.1103/PhysRevLett.131.041002}{{\em Phys. Rev.
  Lett.} {\bfseries 131} no.~4, (2023) 041002},
  \href{http://arxiv.org/abs/2207.03764}{{\ttfamily arXiv:2207.03764
  [hep-ex]}}.

\bibitem{LZCollaboration:2024lux}
{\bfseries LZ Collaboration} Collaboration, J.~Aalbers {\em et~al.}, ``{Dark
  Matter Search Results from 4.2 Tonne-Years of Exposure of the LUX-ZEPLIN (LZ)
  Experiment},'' \href{http://arxiv.org/abs/2410.17036}{{\ttfamily
  arXiv:2410.17036 [hep-ex]}}.

\bibitem{XENON:2023cxc}
{\bfseries XENON} Collaboration, E.~Aprile {\em et~al.}, ``{First Dark Matter
  Search with Nuclear Recoils from the XENONnT Experiment},''
  \href{http://dx.doi.org/10.1103/PhysRevLett.131.041003}{{\em Phys. Rev.
  Lett.} {\bfseries 131} no.~4, (2023) 041003},
  \href{http://arxiv.org/abs/2303.14729}{{\ttfamily arXiv:2303.14729
  [hep-ex]}}.

\bibitem{PandaX:2024qfu}
{\bfseries PandaX} Collaboration, Z.~Bo {\em et~al.}, ``{Dark Matter Search
  Results from 1.54 Tonne$\cdot$Year Exposure of PandaX-4T},''
  \href{http://arxiv.org/abs/2408.00664}{{\ttfamily arXiv:2408.00664
  [hep-ex]}}.

\bibitem{ATLAS:2016wab}
{\bfseries ATLAS} Collaboration, M.~Aaboud {\em et~al.}, ``{Search for scalar
  leptoquarks in pp collisions at $\sqrt{s}$ = 13 TeV with the ATLAS
  experiment},'' \href{http://dx.doi.org/10.1088/1367-2630/18/9/093016}{{\em
  New J. Phys.} {\bfseries 18} no.~9, (2016) 093016},
  \href{http://arxiv.org/abs/1605.06035}{{\ttfamily arXiv:1605.06035
  [hep-ex]}}.

\bibitem{CMS:2017abv}
{\bfseries CMS} Collaboration, A.~M. Sirunyan {\em et~al.}, ``{Search for
  supersymmetry in multijet events with missing transverse momentum in
  proton-proton collisions at 13 TeV},''
  \href{http://dx.doi.org/10.1103/PhysRevD.96.032003}{{\em Phys. Rev. D}
  {\bfseries 96} no.~3, (2017) 032003},
  \href{http://arxiv.org/abs/1704.07781}{{\ttfamily arXiv:1704.07781
  [hep-ex]}}.

\bibitem{ATLAS:2017mjy}
{\bfseries ATLAS} Collaboration, M.~Aaboud {\em et~al.}, ``{Search for squarks
  and gluinos in final states with jets and missing transverse momentum using
  36 fb$^{-1}$ of $\sqrt{s}=13$ TeV pp collision data with the ATLAS
  detector},'' \href{http://dx.doi.org/10.1103/PhysRevD.97.112001}{{\em Phys.
  Rev. D} {\bfseries 97} no.~11, (2018) 112001},
  \href{http://arxiv.org/abs/1712.02332}{{\ttfamily arXiv:1712.02332
  [hep-ex]}}.

\bibitem{CMS:2019zmd}
{\bfseries CMS} Collaboration, T.~C. Collaboration {\em et~al.}, ``{Search for
  supersymmetry in proton-proton collisions at 13 TeV in final states with jets
  and missing transverse momentum},''
  \href{http://dx.doi.org/10.1007/JHEP10(2019)244}{{\em JHEP} {\bfseries 10}
  (2019) 244}, \href{http://arxiv.org/abs/1908.04722}{{\ttfamily
  arXiv:1908.04722 [hep-ex]}}.

\bibitem{Staub:2015kfa}
F.~Staub, ``{Exploring new models in all detail with SARAH},''
  \href{http://dx.doi.org/10.1155/2015/840780}{{\em Adv. High Energy Phys.}
  {\bfseries 2015} (2015) 840780},
  \href{http://arxiv.org/abs/1503.04200}{{\ttfamily arXiv:1503.04200
  [hep-ph]}}.

\bibitem{Alwall:2014hca}
J.~Alwall, R.~Frederix, S.~Frixione, V.~Hirschi, F.~Maltoni, O.~Mattelaer,
  H.~S. Shao, T.~Stelzer, P.~Torrielli, and M.~Zaro, ``{The automated
  computation of tree-level and next-to-leading order differential cross
  sections, and their matching to parton shower simulations},''
  \href{http://dx.doi.org/10.1007/JHEP07(2014)079}{{\em JHEP} {\bfseries 07}
  (2014) 079}, \href{http://arxiv.org/abs/1405.0301}{{\ttfamily arXiv:1405.0301
  [hep-ph]}}.

\bibitem{Dorsner:2016wpm}
I.~Dor\v{s}ner, S.~Fajfer, A.~Greljo, J.~F. Kamenik, and N.~Ko\v{s}nik,
  ``{Physics of leptoquarks in precision experiments and at particle
  colliders},'' \href{http://dx.doi.org/10.1016/j.physrep.2016.06.001}{{\em
  Phys. Rept.} {\bfseries 641} (2016) 1--68},
  \href{http://arxiv.org/abs/1603.04993}{{\ttfamily arXiv:1603.04993
  [hep-ph]}}.

\bibitem{FCC:2018vvp}
{\bfseries FCC} Collaboration, A.~Abada {\em et~al.}, ``{FCC-hh: The Hadron
  Collider}: {Future Circular Collider Conceptual Design Report Volume 3},''
  \href{http://dx.doi.org/10.1140/epjst/e2019-900087-0}{{\em Eur. Phys. J. ST}
  {\bfseries 228} no.~4, (2019) 755--1107}.

\bibitem{Porod:2011nf}
W.~Porod and F.~Staub, ``{SPheno 3.1: Extensions including flavour, CP-phases
  and models beyond the MSSM},''
  \href{http://dx.doi.org/10.1016/j.cpc.2012.05.021}{{\em Comput. Phys.
  Commun.} {\bfseries 183} (2012) 2458--2469},
  \href{http://arxiv.org/abs/1104.1573}{{\ttfamily arXiv:1104.1573 [hep-ph]}}.

\bibitem{Belanger:2014vza}
G.~B\'elanger, F.~Boudjema, A.~Pukhov, and A.~Semenov, ``{micrOMEGAs4.1: two
  dark matter candidates},''
  \href{http://dx.doi.org/10.1016/j.cpc.2015.03.003}{{\em Comput. Phys.
  Commun.} {\bfseries 192} (2015) 322--329},
  \href{http://arxiv.org/abs/1407.6129}{{\ttfamily arXiv:1407.6129 [hep-ph]}}.

\end{thebibliography}\endgroup
\end{document}